\pdfoutput=1
\documentclass[a4paper,11pt]{article}

\usepackage{jheppub} 

\usepackage[T1]{fontenc} 
\usepackage{lmodern}
\usepackage{newtxtext,amssymb,amsmath,extarrows}
\usepackage{graphicx,microtype,physics,framed,empheq}

\newcommand*\widefbox[1]{\fbox{\hspace{2em}#1\hspace{2em}}}

\graphicspath{{fig/}}

\title{\boldmath Cutting Rule for Cosmological Collider Signals: A Bulk Evolution Perspective}

\author{Xi Tong, }
\author{Yi Wang}
\author{and Yuhang Zhu}

\affiliation[]{Department of Physics, The Hong Kong University of Science and Technology,\\Clear Water Bay, Kowloon, Hong Kong, P.R. China}
\affiliation[]{The HKUST Jockey Club Institute for Advanced Study,\\The Hong Kong University of Science and Technology,\\Clear Water Bay, Kowloon, Hong Kong, P.R. China}

\emailAdd{xtongac@connect.ust.hk}
\emailAdd{phyw@ust.hk}
\emailAdd{yzhucc@connect.ust.hk}

\abstract{We show that the evolution of interacting massive particles in the de Sitter bulk can be understood at leading order as a series of resonant decay and production events. From this perspective, we classify the cosmological collider signals into local and nonlocal categories with drastically different physical origins. This further allows us to derive a cutting rule for efficiently extracting these cosmological collider signals in an analytical fashion. Our cutting rule is a practical way for extracting cosmological collider signals in model building, and can be readily implemented as symbolic computational packages in the future.}

\begin{document}
	\maketitle
	\flushbottom
	
	\section{Introduction}\label{Intro}
	As the leading paradigm of the primordial universe, inflation provides us more than a natural genesis of a flat, causal, homogeneous and isotropic background spacetime. Not only can the Gaussian quantum fluctuations prepared by inflation seed the large-scale inhomogeneities and anisotropies, but also the deviations from a purely Gaussian spectrum characterize unique physics during the inflationary era. The study of such non-Gaussianities and its implications for the particle spectrum during inflation has become an active field in the recent years, under the title Cosmological Collider (CC) physics \cite{Chen:2009we,Chen:2009zp,Arkani-Hamed:2015bza,Baumann:2011nk,Assassi:2012zq,Chen:2012ge,Pi:2012gf,Gong:2013sma,Chen:2015lza,Chen:2016nrs,Lee:2016vti,Meerburg:2016zdz,Chen:2016uwp,Chen:2016hrz,Chen:2017ryl,An:2017hlx,An:2017rwo,Kumar:2017ecc,Chen:2018xck,Chua:2018dqh,Kumar:2018jxz,Wu:2018lmx,Alexander:2019vtb,Lu:2019tjj,Hook:2019zxa,Hook:2019vcn,Kumar:2019ebj,Wang:2019gbi,Liu:2019fag,Wang:2019gok,Wang:2020uic,Li:2020xwr,Wang:2020ioa,Fan:2020xgh,Kogai:2020vzz,Bodas:2020yho,Aoki:2020zbj,Arkani-Hamed:2018kmz,Baumann:2019oyu,Baumann:2020dch,Lu:2021gso, Lu:2021wxu,Wang:2021qez}. Powered by the fast expansion of the universe, the CC operates at an energy scale as high as $H\lesssim 10^{14}$GeV, where $H$ is the Hubble parameter during inflation. The inflating spacetime stretches the wavelength of individual modes of a quantum field, leading to the pairwise production of massive particles. If coupled to the inflaton sector, these gravitationally produced particles decay into curvature fluctuations, thereby seeding the non-Gaussian features on the Cosmic Microwave Background (CMB), the Large Scale Structure (LSS), and the 21-cm tomography of neutral hydrogen atoms in the dark age. Considering the coupling to the graviton, these particles also leave distinctive features for the multi-spectra of Primordial Gravitational Waves (PGWs) as well as their cross-correlators with the curvature perturbations. 
	
	More specifically, the observables in cosmological collider physics are the $n$-point correlation functions. In the soft limits, the $n$-point correlation functions exhibit oscillations with respect to the logarithm of momentum ratios ($i.e.$, $\sin(\mu\ln k+\cdots)$), with the frequency $\mu$ encoding the mass of the internal particle $m\sim\mu H$. These oscillations non-analytic in $k$ are the essential ingredient in CC physics, and are often referred to as CC signals or the clock signals~\cite{Arkani-Hamed:2015bza,Chen:2015lza}. Being an on-shell effect, the CC signals are the reminiscence of the mass poles for Mandelstam variables in flat spacetime scattering amplitudes. Furthermore, the angular dependence of the CC signal reveals the spin of the internal particle, while its imaginary part reflects the CP property of the internal particle. 
	
	The past decade has witnessed a rich development of cosmological collider physics. On the theory side, in addition to the traditional in-in formalism \cite{Weinberg:2005vy} and the Schwinger-Keldysh diagrammatics \cite{Chen:2017ryl}, new formalisms are put forward to compute and understand the processes happening at the CC. For instance, based on dS symmetries and the singularity structure of perturbation theory, the cosmological bootstrap program aims to systematically reconstruct the correlation functions from a pure boundary perspective \cite{Arkani-Hamed:2018kmz,Baumann:2019oyu,Baumann:2020dch}. Methods from the AdS toolkit, such as the Mellin-Barnes representation, are applied to tree-level processes with useful results \cite{Sleight:2019mgd,Sleight:2019hfp}. The recent cutting rules utilize locality, unitarity and analyticity, to derive general recursion relations for the discontinuity of cosmological wavefunctions \cite{Goodhew:2020hob,Melville:2021lst,Jazayeri:2021fvk,Goodhew:2021oqg,Baumann:2021fxj,Meltzer:2021zin}. More formally, the cosmological polytopes provide an intriguing geometric picture for the CC process, while also hinting connections between cosmological correlators and flat spacetime scattering amplitudes \cite{Arkani-Hamed:2017fdk}. On the phenomenology side, first, the detection of any Beyond Standard Model (BSM) physics at the CC requires a clear understanding of the Standard Model (SM) background, which has been carefully studied in\cite{Chen:2016uwp,Chen:2016hrz,Kumar:2017ecc,Hook:2019vcn}. BSM models such as supersymmetry \cite{Baumann:2011nk}, extra dimension \cite{Kumar:2018jxz} and monodromy \cite{Flauger:2016idt} give robust predictions at the CC. In search of a large CC signal, CP-violating dimension-5 operators that act as an external chemical potential can assist gravitational production by alleviating the effective mass \cite{Chen:2018xck,Liu:2019fag,Wang:2019gbi,Wang:2020ioa,Hook:2019zxa,Sou:2021juh,Bodas:2020yho}, while a warm inflationary background can provide abundant massive particles from the furnace in the UV \cite{Berera:1995wh,Berera:1995ie,Tong:2018tqf}. As a channel alternative to curvature perturbations, isocurvature colliders also encode the CC signals via modulated reheating \cite{Lu:2019tjj, Kumar:2019ebj, Lu:2021gso}.
	
	However, despite the fruitful developments and immense opportunities for cosmological collider physics, model building in cosmological collider physics still faces a practicality problem. Namely, given a model with a set of couplings, the process of computing CC observables itself can be laborious. Unlike flat spacetime amplitudes, the loss of time translation invariance implies energy non-conservation at interaction vertices, as well as non-trivial mode functions. 
	Due to these two complications, the well-celebrated Lorentz-covariant perturbation theory in the 4-momentum space loses its advantage over the time-dependent perturbation theory in the time domain. However, the time-domain calculation has its own difficulties. A consistent time-dependent perturbation demands specific orderings of the interaction vertices (which, in flat spacetime amplitudes, would be easily handled by $\pm i\epsilon$ deformations in the propagator momenta). This leads to nested integrals too complicated to solve analytically. It is not until the recent developments with the bootstrap method, that a single tree-level exchange diagram is fully evaluated in a closed form. Yet the bootstrap method highly depends on the dS symmetries, which can be spontaneously broken by a rolling scalar background. For example, in the chemical potential scenario, the rolling inflaton selects a preferred dS patch, enhancing dS-symmetry-breaking particle production. So far, CC signals in this scenario have only been evaluated numerically for particles with mass $m\lesssim 5H$, or approximated by using the late-time expansion. However, numerics can be extremely inefficient for large masses, while the late-time expansion does not capture the full waveform of the CC signal, and can even be problematic in certain cases. As a result, how to efficiently extract precise CC signals from the complicated time integrals is a problem faced by all model builders in cosmological collider physics.
	
	We argue that at tree-level, there may be a practical shortcut to solve this problem by considering the physical evolution history in the dS bulk. Although the many nested time integrals obscure the physics, what happens in the bulk at leading order may be described as a combination of three basic events. (i), particles are produced from the background evolution. (ii), particles are produced at interaction vertices. (iii), particles decay into massless inflatons/gravitons. All three types of events happen locally in time, in the form of resonances. Thus the proper time of the particle spent between the resonance events is well defined and corresponds to its dynamical phase. This is the physical origin of the oscillatory CC signals. Hence, depending on the world line of the massive particle, $e.g.$, between (i) and (iii) or between (ii) and (iii), the CC signals can be classified into two categories with different origins and behaviors. On the other hand, these resonances are the stationary points dominating the non-analytic part of the time integral. This motivates us to concentrate on the resonant production/decay picture and propose a practical cutting rule to analytically extract the leading order CC signals for any IR convergent tree diagrams. For the convenience of model builders in cosmological collider physics, we first give a condensed summary of the procedure for this practical cutting rule below and give justifications in the main text.
	\begin{framed}
		{\noindent {\it Cutting rule for CC signals}\quad(For a more detailed version, see Sect.~\ref{CuttingRuleSummary})
			\begin{enumerate}
				\item Given a specific tree diagram, consider only single-colored diagrams with vertices being all-black ($+$) or all-white ($-$).
				\item Focus on one massive propagator at a time and integrate out all other massive propagators as local EFT operators.
				\item Explicitly compute the left/right blobs. The results should be reducible to a sum of polynomials multiplying an exponential function of conformal time. 
				\item Flip the time-ordering Heaviside step function, and obtain two factorized integrals in the form of Laplace transformations. Discard the leftover commutator integral, which contains no CC signals.
				\item Symmetrize and traverse all the left/right injection frequency.
				\item Repeat Step 2-5 for each massive propagator, then sum them up to obtain the total signal.
				\item At last, dress the total signal by the EFT background where all massive fields are integrated out.
			\end{enumerate}
		}
	\end{framed}

	This paper is organized as follows. We first classify the oscillatory CC signals and discuss their physical interpretations from a bulk perspective in Sect.~\ref{ClassificationSect}. Then, in Sect.~\ref{CuttingRuleSect}, we introduce the practical cutting rule for the CC signal and justify it using the resonant production/decay picture. This is supplemented with several examples in Sect.~\ref{ApplicationSect}, including the application to the chemical potential scenario with significant dS symmetry breaking. At last, we conclude and give future prospects in Sect.~\ref{SummarySect}.

	\section{Classification and Interpretation of Cosmological Collider Signals}\label{ClassificationSect}
	\subsection{Cosmological collider physics in a nutshell}\label{CCPReview}
	In this section, we begin with a lightning review of cosmological collider physics, and then introduce the resonant production/decay picture for the bulk evolution of massive particles. During inflation, the inflaton background $\phi_0$ usually evolves into a slow-roll attractor phase with $\dot{\phi}_0\sim$ const. The scale factor in the FRW ansatz grows exponentially with cosmic time, $i.e.$, $a(t)\propto e^{Ht}=-/H\tau$, with $H\sim$ const being the Hubble parameter and $-\infty<\tau<0$ being the conformal time. Thus the background spacetime resembles that of dS space in the Poincaré patch. Upon this quasi-dS background, perturbations in the matter sector can be quantized and calculated. In particular, quantum fluctuations of the inflaton field $\varphi\equiv\phi-\phi_0$ generate the curvature fluctuation $\zeta$ we observed today, via the gauge transformation $\zeta=-(H/\dot{\phi_0})\varphi$. These inflaton perturbations may enjoy non-trivial couplings to various massive particles in the dS bulk, and due to their vanishing mass, they survive the dS expansion until the dS boundary at $\tau=0$. Thus their $n$-point correlation functions $\langle \zeta(\tau_f,\mathbf{x}_1),\dots\zeta(\tau_f,\mathbf{x}_n) \rangle|_{\tau_f\to 0}$ encode interesting information about the particle spectrum of the primordial universe. 
	
	To calculate the correlation functions in a given particle physics model with weak couplings, it is most straightforward to use a diagrammatic representation to organize the perturbative expansion. This formalism is known as the Schwinger-Keldysh path integral. The quantum fields start out in the Bunch-Davis (BD) vacuum at $\tau_i\to-\infty$, then evolve according to a local Hamiltonian density, and are finally collected by the $n$ probes placed at $\tau_f\to 0$. We read out the \textit{average value} of the $n$ localized probes instead of the \textit{probability amplitudes} of detecting certain on-shell states which are non-local in spacetime. Due to this special setup, only the ``in'' state is well-defined but not the ``out'' state. We also need two copies of evolution histories to accomplish the averaging process, one with time ordering ($+$) and the other with anti-time ordering ($-$). More specifically, take the inflaton as the example, the generating functional of non-interacting curvature fluctuations can be written as
	\begin{equation}
	Z_0[J_+,J_-]\equiv\int\mathcal{D}\zeta_+\mathcal{D}\zeta_-\text{exp}\left[i\int_{\tau_i}^{\tau_f}d\tau d^3x\left(\mathcal{L}_0[\zeta_+]-\mathcal{L}_0[\zeta_-]+J_+\zeta_+-J_-\zeta_-\right)\right].
	\end{equation}
	There are totally four types of propagators which can be generated using
	\begin{align}
	-iG_{ab}(\tau,\mathbf{x_1};\tau',\mathbf{x_2})=\frac{\delta}{ia\delta J_a(\tau,\mathbf{x_1})}\frac{\delta}{ib\delta J_b(\tau',\mathbf{x_2})}Z_0[J_+,J_-]\bigg|_{J_{\pm}=0}~,~  a,b=\pm~.	
	\end{align}
	In momentum space, they are
	\begin{eqnarray}
	G_{++}(\mathbf{k},\tau,\tau')&=&\theta(\tau-\tau')u_k(\tau)u^*_k(\tau')+\theta(\tau'-\tau)u^*_k(\tau)u_k(\tau')~,\\
	G_{+-}(\mathbf{k},\tau,\tau')&=&u^*_k(\tau)u_k(\tau')~,\\
	G_{-+}(\mathbf{k},\tau,\tau')&=&u_k(\tau)u^*_k(\tau')~,\\
	G_{--}(\mathbf{k},\tau,\tau')&=&\theta(\tau-\tau')u^*_k(\tau)u_k(\tau')+\theta(\tau'-\tau)u_k(\tau)u^*_k(\tau')~,
	\end{eqnarray}
	where $u_k(\tau)$ is the inflaton mode function. The propagator $D_{ab}$ for massive fields can be derived in a similar way, with $u_k(\tau)$ replaced by their own mode functions $v_k(\tau)$. As a reference, under the BD initial condition, the scalar mode function reads
	\begin{align}
	& u_k=\frac{H}{2\sqrt{\epsilon}{M_{\rm{P}}}}\frac{1}{k^{3/2}}(1+ik\tau)e^{-ik\tau}~,\\
	&v_k=-ie^{-\frac{\pi}{2}\mu+i\frac{\pi}{4}}\frac{\sqrt\pi}{2}H(-\tau)^{3/2}H^{(1)}_{i\mu}(-k\tau)~,~~\mu\equiv\sqrt{\frac{m^2}{H^2}-\frac{9}{4}}~.
	\end{align}
	
    \begin{figure}[htp] 
	\centering
    \includegraphics[width=15cm]{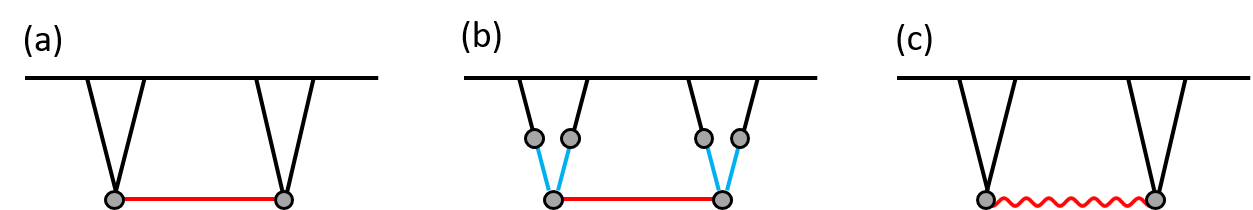}
	\caption{\label{typicaldiam} Some typical diagrams one encounters in cosmological collider physics. They are 4-point diagrams with (a) an internal massive scalar propagator (red line), (b) an internal massive propagator (red line) and four massive propagators of another mass (blue line) on the external legs, (c) a massive vector propagator (red wavy line).}
    \end{figure}
    Interaction terms can be perturbatively expanded as usual. After separating the Lagrangian into two parts,
    \begin{equation}
    \mathcal{L}=\mathcal{L}_0+\mathcal{L}_{\rm{int}}~,
    \end{equation}
    the generating functional can be written as
    \begin{align}
    Z[J_+,J_-]=\exp\left[i\int_{\tau_i}^{\tau_f}d\tau d^3x\left(\mathcal{L}_{\rm{int}}\left[\frac{\delta}{i \delta  J_+}\right]-\mathcal{L}_{\rm{int}}\left[-\frac{\delta}{i\delta J_- }\right]\right)\right]Z_0\left[J_+,J_-\right].
    \end{align}
    For weak coupling sizes, we are able to use time-dependent perturbation theory and expand the above functional to any desired order. More specifically, for a given diagram, each internal vertex may be colored to indicate their time ordering. Black vertices ($+$) represent time ordering while white vertices ($-$) represent anti-time ordering. 
    For a diagram with $V$ vertices, we must sum the $2^V$ colored diagrams to obtain the final result. For instance, Fig.~\ref{typicaldiam} shows three typical diagrams one may encounter in cosmological collider physics.
    
    As stated in Sect.~\ref{Intro}, the reason why correlation functions in dS are difficult to compute is that the time-domain calculation requires doing nested integrals of special functions. Take diagram (b) in Fig.~\ref{typicaldiam}, for example, there are effectively $2^6/2=32$ colored diagrams for us to calculate\footnote{Complex conjugation reduces the number of colored diagram by a factor of two.}. Of these 32 diagrams, one contains a $6$-layer time integral, four contain a $5$-layer integral, and six contain a $4$-layer integral, etc. Considering also that the integrand consists of products of ten Hankel/Whittaker functions, it seems rather hard to extract the desired analytical form of CC signals. With the help of effective propagators introduced in \cite{Chen:2017ryl}, one can numerically evaluate this diagram for small masses. However, in the presence of chemical potential, if the mass of the internal particles is large ($e.g.$, $m\gtrsim 5H$), memory consumption quickly becomes unmanageable due to slow integral convergence. Moreover, the effective propagator is limited to handling two-point mixing between massive field and the external inflaton only, and is not applicable to more general couplings. Therefore, we see that at tree level already, we are in need of a more efficient way to extract CC signals and physics.

	\subsection{Bulk evolution as resonance events}
	To find an efficient way to extract the CC signals, we should understand the evolution history of massive fields in the dS bulk. First, let us consider a free massive field. Due to the curved background geometry, the mode function of massive field usually takes the form of special functions ($e.g.$ Hankel or Whitaker functions). Yet the mathematical complexity of special functions may conceal the physics inside. Alternatively, a more physical approach is to use the super-adiabatic approximation based on the Stokes-line method~\cite{Sou:2021juh}.
	Consider the equation of motion of a canonically normalized massive field with mode function $v_{k_I}(\tau)$,
	\begin{align}
	\frac{d^2}{d\tau^2}v_{k_I}(\tau)+w^2(\tau)v_{k_I}(\tau)=0~,\qquad
	w^2=k_I^2+m^2 a^2-\frac{a''}{ a}~.
	\end{align} 
    The solution in terms of the super-adiabatic approximation reads
	\begin{equation}\label{sabm}
    v_{k_I}(\tau)=\alpha(\tau) f(\tau)+\beta(\tau) f^*(\tau),
	\end{equation}
	\begin{equation}
	\text{with}~~f(\tau)\equiv\frac{1}{\sqrt{2w(\tau)}}e^{-i\int_{\tau_i}^{\tau}w(\tau')d\tau'},~ \alpha(\tau)\approx 1~~ \text{and}~~\beta(\tau)\approx -iS(\tau)e^{-F(\tau_i)}~.
	\end{equation}
	Here $S(\tau)$ is the Stokes multiplier and $F(z)$ is Dingle's singulant variable,
	\begin{align}
	S(\tau)=\frac{1}{2}\left[1+\text{Erf}\left(\frac{-\Im F(\tau)}{\sqrt{2|\Re F(\tau)|}}\right)\right],\qquad F(\tau)=2i\int_{\tau_c}^{\tau}w(\tau_1)d\tau_1~,
	\end{align}
	and $\tau_c$ is the complex turning point in the lower-half complex plane, $i.e.$, $w^2(\tau_c)=0,\Im \tau_c<0$. The above division between the positive frequency part $f(\tau)$ and negative frequency part $f^*(\tau)$ minimizes the oscillations in $\alpha$ and $\beta$ by truncating the super-adiabatic series at an optimal order \cite{Dabrowski:2016tsx}. This results in the most natural choice of a time-dependent particle number,
	\begin{equation}
		\langle n_{k_I}(\tau)\rangle'=|\beta(\tau)|^2\approx e^{-2\Re F(\tau_i)}S^2(\tau)~.
	\end{equation}
	Therefore, it is clear that gravitational particle production happens when $\Im F(\tau_*)=0$, at
	\begin{equation}
		\textbf{Production time:}\qquad\qquad |k_I\tau_*|\approx0.66\mu~,\qquad\qquad\qquad\qquad\quad
	\end{equation}
	with a duration 
	\begin{equation}
		k_I\Delta \tau_*=\frac{2\sqrt{2|\Re F(\tau_*)|}}{|\Im F'(\tau_*)|}\approx 1.4\sqrt{\mu}~.
	\end{equation}
	After production, the initial BD vacuum evolves into a two-particle squeezed state via the Bogoliubov transformation defined by (\ref{sabm}),
	\begin{equation}
		|\text{BD}\rangle
		= |0\rangle+ \frac{\beta}{2} \int_\mathbf{k} |1_\mathbf{k}\rangle\otimes|1_{-\mathbf{k}}\rangle+\mathcal{O}(\beta^2)~.
	\end{equation}
	For $|\Delta \tau_*/\tau_*|\lesssim 1$, which is typical for massive fields, the production of the particle pair can be viewed as an event localized in time. $\tau_*$ thus marks the starting tick of the clock measured by the dynamical phase of the massive particle. Upon production, the particle pair will propagate in opposite directions and accumulate their dynamical phases in the process. This can be seen more clearly in the Wightman function,
	\begin{equation}
		\langle\sigma_-(\tau,\mathbf{k}_I)\sigma_+(\tau',\mathbf{-k}_I)\rangle=D_{-+}(\mathbf{k}_I,\tau,\tau')=D_{-+}^{\text{non-local}}(\mathbf{k}_I,\tau,\tau')+D_{-+}^{\text{local}}(\mathbf{k}_I,\tau,\tau')~,
	\end{equation}
	with
	\begin{eqnarray}
		D_{-+}^{\text{non-local}}(\mathbf{k}_I,\tau,\tau')&=&\alpha(\tau)\beta^*(\tau')f_{k_I}(\tau)f_{k_I}(\tau')+\alpha^*(\tau')\beta(\tau) f^*_{k_I}(\tau)f^*_{k_I}(\tau')~,\\
		D_{-+}^{\text{local}}(\mathbf{k}_I,\tau,\tau')&=&\alpha(\tau)\alpha^*(\tau')f_{k_I}(\tau)f^*_{k_I}(\tau')+\beta(\tau)\beta^*(\tau') f^*_{k_I}(\tau)f_{k_I}(\tau')~.
	\end{eqnarray}
	The late-time limit ($\tau,\tau'\to 0$) momentum dependence defines the local vs non-local separation\footnote{Another way of this division is in terms of Bessel functions~\cite{Lee:2016vti}, $i.e.$, $J_{i\mu}(-k\tau)$ and $J_{-i\mu}^*(-k\tau)$. Nevertheless, this division between the local and non-local part is not as sharp. Because before the production time, the negative frequency part has not appeared yet and non-local part propagator should still stay at zero value.},
	\begin{align}
	\nonumber\underset{\tau,\tau'\rightarrow 0}{\rm{lim}}D_{-+}^{\text{non-local}}(\mathbf{k}_I,\tau,\tau')&~=~ \frac{H^2(\tau\tau')^{3/2}}{4\pi}\left[\Gamma(-i\mu)^2\left(\frac{{k_I}^2\tau\tau'}{4}\right)^{i\mu}+\Gamma(i\mu)^2\left(\frac{{k_I}^2\tau\tau'}{4}\right)^{-i\mu}\right]~,\\
	\underset{\tau,\tau'\rightarrow 0}{\rm{lim}}D_{-+}^{\text{local}}(\mathbf{k}_I,\tau,\tau')&~=~ \frac{H^2(\tau\tau')^{3/2}}{4\pi}\Gamma(-i\mu)\Gamma(i\mu)\left[e^{\pi\mu}\left(\frac{\tau}{\tau'}\right)^{i\mu}+e^{-\pi\mu}\left(\frac{\tau}{\tau'}\right)^{-i\mu}\right].\label{MassivePropIR}
	\end{align}
	The local part is a constant independent of $k_I$ and upon Fourier transformation back to space coordinate, gives a contact term proportional to $\delta^3(\mathbf{x}_1-\mathbf{x}_2)$, hence the locality. The non-local part is non-analytic in $k_I$, with the $k_I^{2i\mu}$ term giving rise to long-range correlations of the form $|\mathbf{x}_1-\mathbf{x}_2|^{2i\mu}$. The non-local part's time dependence manifests the accumulated dynamical phase of \textit{two} propagating particles,
	\begin{equation}
		(k_I^2\tau\tau')^{\pm i\mu}\propto \left(\frac{\tau}{\tau_*}\right)^{\pm i\mu}\left(\frac{\tau'}{\tau_*}\right)^{\pm i\mu} \sim e^{\mp im(t-t_*)}e^{\mp i m(t'-t_*)}~,
	\end{equation}
	where $t$ stands for the corresponding cosmic time.	In contrast, the local part's time dependence is of the form $\left(\tau/\tau'\right)^{\pm i\mu}$. This can be viewed as the dynamical phase of \textit{one} single particle propagating from $\tau'$ to $\tau$ (or $\tau$ to $\tau'$, but with an extra $e^{-2\pi\mu}$ suppression):
	\begin{equation}
		\left(\frac{\tau}{\tau'}\right)^{\pm i\mu}\sim e^{\mp i m(t-t')}~.
	\end{equation}
	Up till now, $\tau,\tau'$ can be any time later than the starting tick at $\tau_*$. We will see later that the introduction of interactions automatically leads to events localized in time. These interactions set the interaction time $\tau,\tau'$ as starting/ending ticks. Before that, we notice a nice property about the non-local part of massive propagator, that it is real after pair production time.
	\begin{align}
	D_{-+}^{\text{non-local}}(\mathbf{k}_I,\tau,\tau')=	D_{-+}^{\text{non-local}}(\mathbf{k}_I,\tau,\tau')^*~,~~\text{with}~~\tau,\tau'>\tau_*~.
	\end{align}
	Consequently, the commutator of field operators does not contain non-local parts:
	\begin{align}\label{nonlocalpro}
	\langle\left[\sigma_{k_I}(\tau),\sigma_{k_I}(\tau')\right]\rangle\supset\left[v^*_{k_I}(\tau)v_{k_I}(\tau')-v_{k_I}(\tau)v^*_{k_I}(\tau')\right]_{\text{from non-local }}=0~.
	\end{align}
	This fact is essentially due to the requirement of microcausality. If the commutator has non-zero non-local parts with $k_I^{\pm i\mu}$ dependence, after Fourier transforming back to position space, the commutator will not vanish out of the light-cone, in direct contradiction with microcausality. The fact above has a direct consequence in the CC signals as we shall see later.
	
\begin{figure}[htp] 
		\centering
		\includegraphics[width=15.5cm]{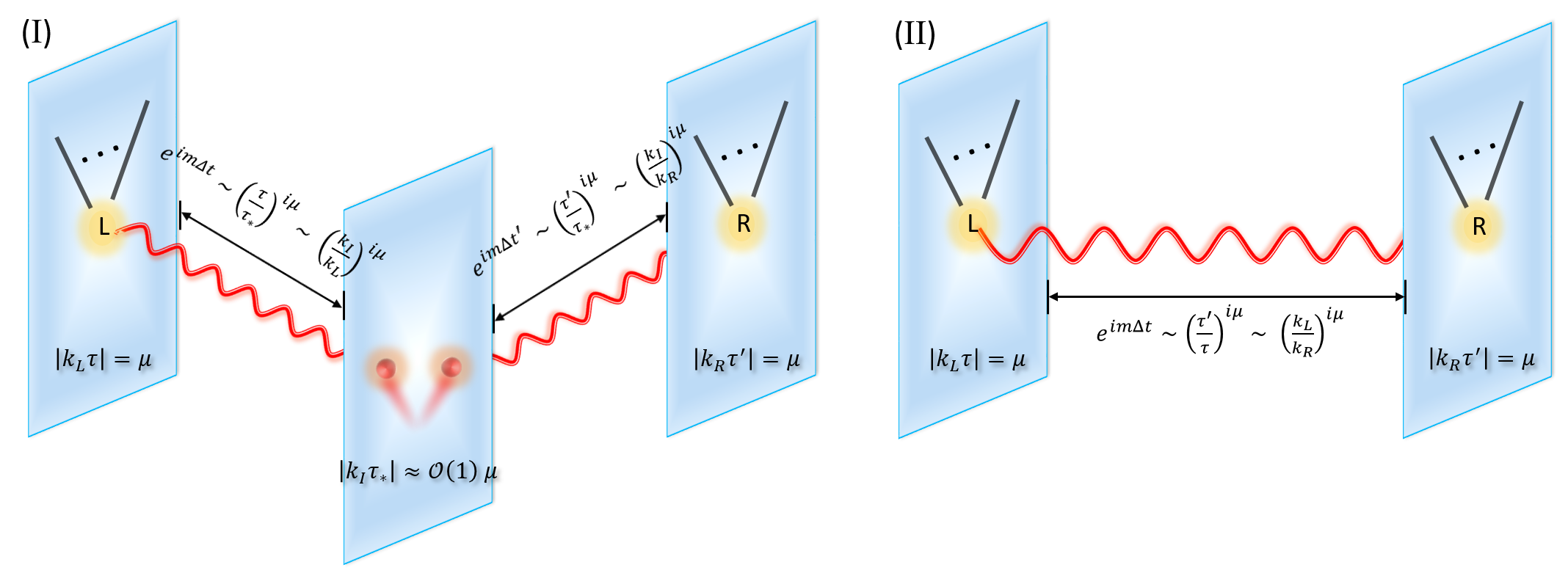}
		\caption{\label{figure1}Illustrations of the physical interpretation of two different types of CC signals, (I) shows the non-local type and (II) is the local type. The planes indicate events happening at different locations in spacetime. In both cases, the dynamical phases accumulated between the events are encoded as oscillations in momentum ratios.}
	\end{figure}

	Now let us turn on the interactions of the massive field and the massless inflaton $\varphi$. Any tree diagram containing a massive field propagator can then be summarized into the form shown in Fig.~\ref{figure1}. The left and right blob represent the fully evaluated subdiagram on the left and right, and at time $\tau,\tau'$, respectively. At early times, $\tau,\tau'\to-\infty$, the left (right) blob has a leading time dependence $e^{i k_L\tau}$ ($e^{i k_R\tau}$), with $k_L$ ($k_R$) containing various momentum combinations from the left (right) subdiagram. For a soft internal momentum $k_I\ll k_L,k_R$, the massive propagator takes the form (\ref{MassivePropIR}), and the integrals over $\tau,\tau'$ pick up the stationary points\footnote{A more accurate calculation of the stationary points can be obtained via the super-adiabatic mode function, which gives $|\sqrt{k_{L}^2-k^2_I}\tau_\bullet|,|\sqrt{k_{R}^2-k^2_I}\tau'_\bullet|\approx\mu$. Usually, in the soft limit where $k_I\ll k_{L},k_R$, the resonance time reduces to (\ref{rescd1}).}
	\begin{equation}
		\int d\tau \tau^{\alpha_L}e^{i k_L\tau}\tau^{\pm i\mu}\sim \frac{\sqrt{2\pi\mu}}{k_L}e^{\pm i\frac{\pi}{4}} \tau_\bullet^{\alpha_L}e^{i k_L\tau_\bullet}\tau_\bullet^{\pm i\mu}
	\end{equation}
	at
	\begin{align}\label{rescd1}
		\textbf{Resonance time:}\qquad\qquad&|k_{L}\tau_{\bullet}|\approx\mu~~~,~~~|k_{R}\tau'_\bullet|\approx\mu~.\qquad\qquad\qquad&
	\end{align}
	These resonances also happen as events localized in time. Physically they can be understood as the decay of the massive particle into massless modes, or its resonant production at the vertex from massless modes. These events effectively mark the ending ticks at $\tau_\bullet,\tau'_\bullet$. 
	
	As a result, we have three types of events happening in the dS bulk. Namely, non-local gravitational production, local vertex resonant production, and local vertex resonant decay. Therefore the massive particle's dynamical phases are accumulated either (\textbf{I}) from the gravitational production event at $\tau_*$ to the resonant decay event at $\tau_\bullet,\tau'_\bullet$, or \textbf{(II}) from the resonance production event at $\tau'_\bullet$ to the resonant decay event at $\tau_\bullet$ (or vice versa). From this classification criterion, we can deduce two possible types of CC signals,
	\begin{empheq}[box=\widefbox]{align}
		\nonumber\textbf{I. }\textbf{Non-local Type:}&\quad \left(\frac{k_L k_R}{k_I^2}\right)^{\mp i\mu}\sim\left(\frac{\tau_\bullet}{\tau_*}\right)^{\pm i\mu}\left(\frac{\tau_\bullet'}{\tau_*}\right)^{\pm i\mu}&\subset&~D_{-+}^{\text{non-local}}(\mathbf{k}_I,\tau_\bullet,\tau'_\bullet) ~,\\
		\nonumber\textbf{II. }\textbf{Local Type:}&\quad \left(\frac{k_L}{k_R}\right)^{\mp i\mu}\sim\left(\frac{\tau_\bullet}{\tau_\bullet'}\right)^{\pm i\mu}&\subset&~D_{-+}^{\text{local}}(\mathbf{k}_I,\tau_\bullet,\tau'_\bullet)~.
	\end{empheq}
	This physical picture of bulk evolution is illustrated in Fig.~\ref{figure1}. According to their different origins (dependence about soft momentum), we will name the signals as the non-local type or the local type. 
	
	Before concluding this section, we make a few remarks on our classification of CC signals.
	\begin{enumerate}
		\item Mathematically speaking, these two categories of CC signals are indeed independent features. Because scale invariant quantities only depend on a subspace in the 3-dimensional parameter space $\mathbb{R}^3$ spanned by $k_L,k_R,k_I$, namely the projective plane $\mathbb{R}P^2$. One can freely choose $k_L k_R/k_I^2$ and $k_L/k_R$ to parametrize two orthogonal directions in this scale-invariant subspace $\mathbb{R}P^2$.
		\item Physically speaking, these two categories of CC signals have drastically different physical origins. As mentioned above, the non-local signal comes from the gravitational production and vertex decay of \textit{two} massive particles, whereas the local signal comes from the vertex production and vertex decay of \textit{one} massive particle. 
		\item Following the above argument, the two CC signals can have different strengths, because of their intrinsically distinct production mechanisms ($i.e.$, one from background at a linear level, one from perturbation at a non-linear level). The non-local part $D_{-+}^{\text{non-local}}$ is suppressed by $|\alpha\beta^*|\sim\sqrt{\langle n_{\mathbf{k}_I}\rangle'}$, while the local part $D_{-+}^{\text{local}}$ is, to leading order, independent of the particle number $\langle n_{\mathbf{k}_I}\rangle'$. This is most clearly demonstrated in Sect.~\ref{ChemPtlSect} in the presence of a chemical potential, which can alter the particle number. We will see that the non-local CC signal strength can be greatly influenced by the chemical potential, whereas the local CC signal strength is, to leading order, fixed by dS geometry alone.
		\item The resonance picture also has the advantage that it always picks up the leading order contribution, since they represent the stationary points in the time integral. If the stationary point lies in the unphysical region ($\tau_\bullet>0$), there is an overall suppression $(-1)^{i\mu}=e^{-\pi\mu}$, which is the penalty for violating energy conservation in dS.\label{suppressionOrigin} 
		\item Notice that in the left and right blobs, the particles need not to be massless. At early times (or equivalently, with hard momenta), all mode functions approach to that of a massless particle with a dynamical phase $e^{-ik\tau}$. Therefore, the intermediate massive particle can resonate with \textit{all} possible frequencies in the left/right blob, giving rise to various patterns of CC signals\footnote{In cases with two-point mixing, the early-time limit of the left/right blobs may be in conflict with the late-time limit of the massive propagator. However, one can still obtain the $e^{ik\tau}$ dynamical phase by integrating out the massive fields in the left/right blobs as EFT operators of massless fields. In fact, this will be our general strategy to be mentioned in Sect.~\ref{npts}.}. Then the physical interpretation is that the massive particle in consideration can be produced by or decay into all possible particle combinations in the left/right blobs, which are still in the UV ``massless'' phase.
		\item Most of the past studies on CC signals focus on the non-local type, with a few exceptions that explicitly mentioned the local type signal \cite{Wang:2020ioa,McAneny:2019epy}. We point out that local signals themselves are not newly revealed result, yet their distinction and interpretation is something we wish to emphasize and exploit in this work.
	\end{enumerate}

	\section{A Cutting Rule for CC Signals}\label{CuttingRuleSect}
	Based on the above observation on bulk evolution, in this section, we propose a practical cutting rule for the efficient extraction of both types of CC signals.
	\subsection{The general method}
	\begin{figure}[htp] 
		\centering
		\includegraphics[width=11cm]{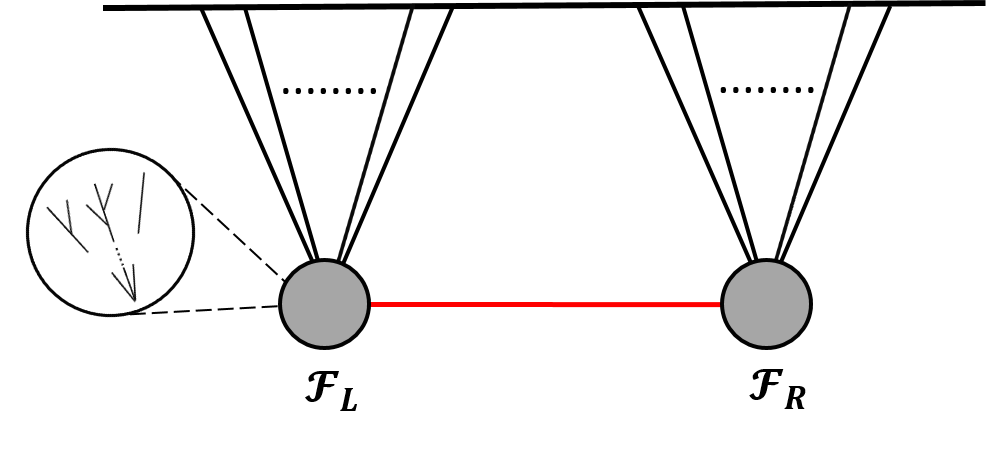}
		\caption{\label{tree} A general tree level diagram. The red line is the massive propagator and other black lines are massless. The left and right blobs may have their own substructure and we collect it into $\mathcal{F}_L$ and $\mathcal{F}_R$~.}
	\end{figure}
	Consider a general $n$-point tree-level diagram (see Fig.~\ref{tree}) with one massive propagator in the Schwinger-Keldysh formalism. Focusing on the internal massive propagator, the diagram can be expressed as the following form,
	\begin{align}
		\langle\zeta^n\rangle\supset\sum_{a,b=\pm}\int_{-\infty}^{0}d\tau d\tau'\mathcal{F}^a_L(\tau)\mathcal{F}^b_R(\tau')D_{ab}(\mathbf{k}_I,\tau,\tau'),
	\end{align}
	where we have absorbed terms related to the left blob and right blob into $\mathcal{F}_L$ and $\mathcal{F}_R$. In general, they consist of products of the scale factor, coupling constants, as well as the detailed expression of the substructures. We also assume that the time integrals inside the substructures have been performed, so that the only time dependences come from the two endpoints ($\tau,\tau'$), to which the massive propagator is attached. In the early-time limit $\tau\to-\infty$, or in the soft limit $k_I\to 0$ (namely, when the massive mode momentum is much smaller than any momentum combinations in the left/right blob) the time dependence of the left/right blob becomes simple:
	\begin{equation}
		\mathcal{F}_{L,R}(\tau)\sim e^{\pm ik_{L,R}\tau}~,
	\end{equation}
	where $k_{L,R}$ represents various combinations of momentum magnitude sums flowing into the internal massive propagator.
	
	Depending on the vertex coloring, the integrand can be separated into the time-ordered part (TO, $ab=++$), the anti-time-ordered part (ATO, $ab=--$) and the non-time-ordered part (NTO, $ab=+-$ or $+-$). The NTO part is easier to deal with, because these two time integrals are factorized and can be performed independently. In other words, diagrams with opposite coloring on any adjacent pair of vertices are already cut and become factorized. What actually complicates the calculation is the nested time integral from the TO part\footnote{The ATO part is obtained from the TO part via complex conjugation and momentum reversal.}, which has bounded integrals such as $\int_{\tau}^{0}$ or $\int_{-\infty}^{\tau}$. However, we find that if one focus on the oscillatory CC signals, which is the most salient feature of cosmological collider physics, then after re-organizing the integrand, one can cut through the time-ordering, and factorize the whole diagram. Consequently, the TO part becomes as simple as the NTO part.
	
	To demonstrate the idea, we focus on the TO part,
	\begin{align}
		\nonumber\langle\zeta^n\rangle_{\rm{TO}}\sim&\int_{-\infty}^{0}d\tau d\tau'\mathcal{F}_L(\tau)\mathcal{F}_R(\tau')
		\\&\times \left[\theta(\tau-\tau')v_{k_I}(\tau)v^*_{k_I}(\tau')+\theta(\tau'-\tau)v^*_{k_I}(\tau)v_{k_I}(\tau')\right]~.
	\end{align} 
	Without loss of generality, let us assume $k_L>k_R$. We can re-organize the integral by flipping one of the Heaviside step function,
	\begin{equation}
		\theta(\tau-\tau')=1-\theta(\tau'-\tau)~,\label{FlippingTheta}
	\end{equation}
	after which the integration is split into two terms,
	\begin{align}
		\nonumber\langle\zeta^n\rangle_{\rm{TO}}\sim&\int_{-\infty}^{0}d\tau \mathcal{F}_L(\tau)v_{k_I}(\tau)\int_{-\infty}^{0}d\tau' \mathcal{F}_R(\tau')v_{k_I}^*(\tau')
		\\&\label{cmint}+\int_{-\infty}^{0}d\tau\int_{\tau}^{0}d\tau'\mathcal{F}_L(\tau)\mathcal{F}_R(\tau')\left[v^*_{k_I}(\tau)v_{k_I}(\tau')-v_{k_I}(\tau)v^*_{k_I}(\tau')\right]~.
	\end{align}
	The first term factorizes into a product of two integrals over $(-\infty,0)$, resembling the easier NTO part. Then we only need to examine the second term with bounded integral, which is essentially the field commutator mentioned in Sect.~\ref{ClassificationSect}. We will argue that this second term is irrelevant for both types of CC signal if $k_L>k_R$.
	
	First of all, it is easy to check that this commutator nevertheless cannot contribute to non-local type CC signals. By using the identity (\ref{nonlocalpro}), the commutator from non-local part is exactly canceled for $\tau'>\tau>\tau_*$ and for $\tau_*>\tau'>\tau$. The only non-zero contribution is with the hierarchy
	\begin{equation}
		\tau'>\tau_*>\tau~.\label{nonzerohierarchy}
	\end{equation}
	Recall that the non-local signal requires a pair production event at $\tau_*$ and two resonant decay events at $\tau_\bullet,\tau'_\bullet$. Thus we must require
	\begin{equation}
		\tau=\tau_\bullet>\tau_*~.\label{prodconstraint}
	\end{equation}
	However, (\ref{nonzerohierarchy}) and (\ref{prodconstraint}) cannot be satisfied simultaneously, suggesting the absence of the non-local type CC signal.
	
	Then the remaining question is whether this commutator contributes to the local type signal. According to our previous physical picture of bulk evolution, the local CC signal can be thought of as the dynamical phase between two resonance events happening at different vertices. The resonance condition at the vertices gives
	\begin{equation}
		|k_L \tau_\bullet|=\mu\qquad\text{and}\qquad|k_R \tau'_\bullet|=\mu~.\label{resonbanceconstraint}
	\end{equation}
	On the other hand, the integration region constrains
	\begin{align}
		k_L>k_R~\qquad\text{and}\qquad|\tau_\bullet|>|\tau'_\bullet|~,\label{timeorderconstraint}
	\end{align}
	However, (\ref{resonbanceconstraint}) and (\ref{timeorderconstraint}) cannot be satisfied simultaneously. Therefore, the local type CC signals are also absent in the commutator integral.
	
	\begin{figure}[tp] 
		\centering
		\includegraphics[width=15cm]{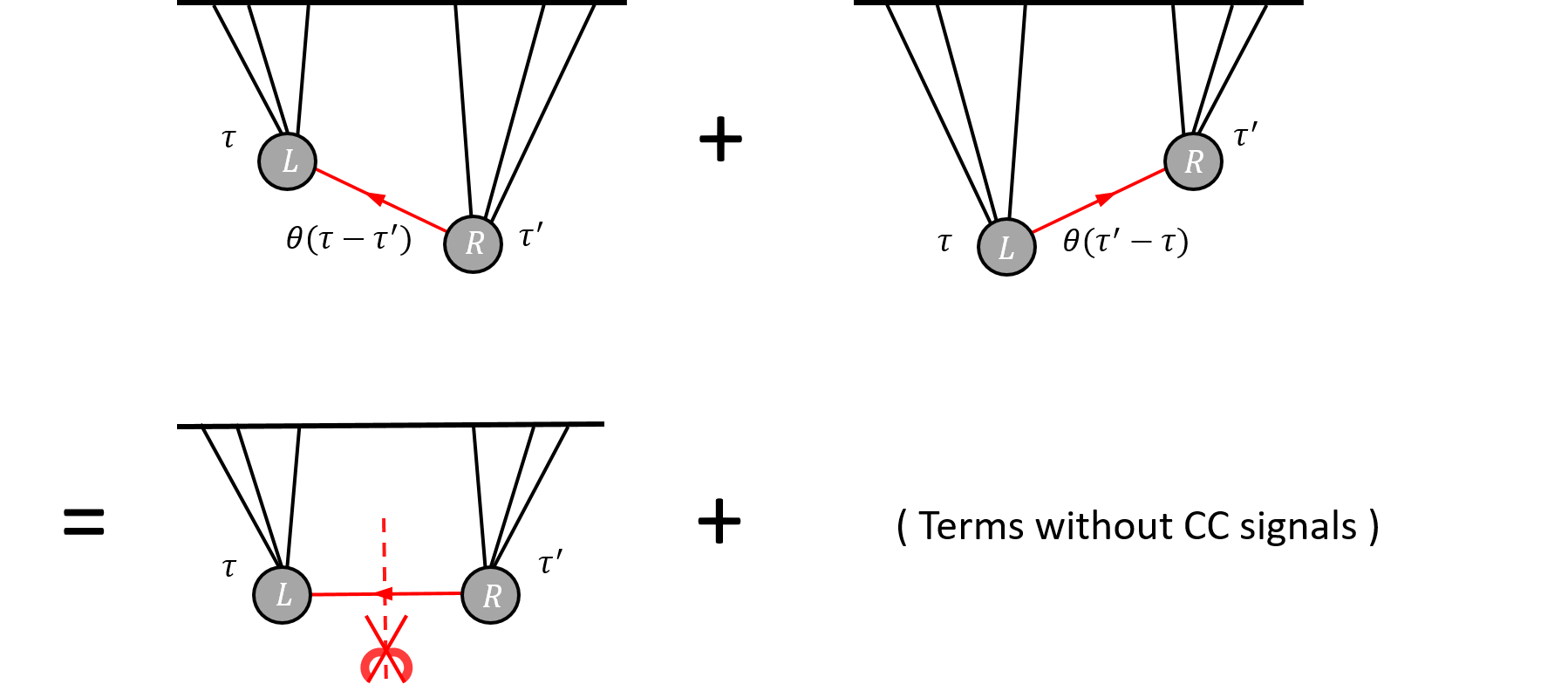}
		\caption{A schematic diagram showing how to cut the nested integral and extract the CC signals. The first row is the TO diagram before the cut. The second row shows a factorized diagram (the cut result) and a commutator integral, which does not contain the CC signals and can be discarded.}\label{cuttingrule}
	\end{figure}
	
	As summarized in Fig.~\ref{cuttingrule}, we have obtained a method to cut the TO integral into factorized integrals, with a small leftover piece free of any desired signals.  Note that our initial assumption of $k_L>k_R$ is crucial for us to argue against the presence of local signals in the leftover commutator integral. However, this is not a limitation to our method. The case with $k_L<k_R$ is completely analogous. Namely, we only need to flip the other Heaviside step function\footnote{Alternatively, one can obtain the cut result in the parameter region $k_L>k_R$ and then perform a symmetrization $k_L\leftrightarrow k_R$.} in (\ref{FlippingTheta}), and the rest of the argument carries over. The complete cut result is then necessarily a \textit{piecewise} function of $k_L$ and $k_R$.

	Before going forward, we can already have a rough estimation of the signal strength using the cut result. After the cut, the signal $S^>_I$ in the TO part can be estimated as
	\begin{eqnarray}
		\nonumber S^>_I
		&\sim&\int_{-\infty}^{0}d\tau \mathcal{F}_L(\tau)\left[\alpha f_{k_I}(\tau)+\beta f^*_{k_I}(\tau)\right]\int_{-\infty}^{0}d\tau' \mathcal{F}_R(\tau')\left[\alpha^* f^*_{k_I}(\tau')+\beta^* f_{k_I}(\tau')\right]\\\nonumber
		&\sim&\mathcal{O}(e^{-\pi\mu})|\alpha|^2\left(\frac{k_{R}}{k_{L}}\right)^{i\mu}+\mathcal{O}(e^{-\pi\mu})|\beta|^2\left(\frac{k_{R}}{k_{L}}\right)^{-i\mu}\\
		&+&\mathcal{O}(1)\alpha\beta^*\left(\frac{k_{L}k_{R}}{k_I^2}\right)^{-i\mu}+\mathcal{O}(e^{-2\pi\mu})\alpha^*\beta\left(\frac{k_{L}k_{R}}{k_I^2}\right)^{i\mu}~.
	\end{eqnarray}
	Clearly, the two types of signals are suppressed by different factors. The non-local signal is due to gravitational particle production at a linear level, hence it is suppressed by $\beta\sim\sqrt{\langle n\rangle'}$. In contrast, the local signal comes from vertex particle production at a non-linear level, thus it is suppressed by the dS intrinsic Boltzmann factor $e^{-\pi\mu}\sim \sqrt{e^{-m/T_{dS}}}$. This is again the penalty of violating energy conservation in dS as mentioned in Point \ref{suppressionOrigin} of the summary above. If the gravitational particle production respects dS symmetries, $\beta=e^{-\pi\mu}$ and the two signal strengths are degenerate. However, with dS-symmetry breaking chemical potential, the particle number can deviate from that of a simple Boltzmann form, leading to the lifting of the signal strength degeneracy. This is explicitly demonstrated with a vector field later in Sect.~\ref{ChemPtlSect}.
	
	As a side remark, it is easy to check the signals in the NTO part,
	\begin{eqnarray}
		\nonumber\mathcal\langle\zeta^4\rangle'_{NTO}
		&\sim&\int_{-\infty}^{0}d\tau \mathcal{F}_L(\tau)\left[\alpha^* f_{k_I}^*(\tau)+\beta^* f_{k_I}(\tau)\right]\int_{-\infty}^{0}d\tau' \mathcal{F}^*_R(\tau')\left[\alpha f_{k_I}(\tau')+\beta f^*_{k_I}(\tau')\right]\\\nonumber
		&\sim&\mathcal{O}(e^{-2\pi\mu})|\alpha|^2\left(\frac{k_{R}}{k_{L}}\right)^{-i\mu}+|\beta|^2\left(\frac{k_{R}}{k_{L}}\right)^{i\mu}\\
		&+&\mathcal{O}(e^{-\pi\mu})\alpha\beta^*\left(\frac{k_{L}k_{R}}{k_I^2}\right)^{-i\mu}+\mathcal{O}(e^{-\pi\mu})\alpha^*\beta\left(\frac{k_{L}k_{R}}{k_I^2}\right)^{i\mu}~.
	\end{eqnarray}
	Obviously, the non-time-order part is heavily suppressed by $|\beta|^2,e^{-\pi\mu}|\beta|$ or $e^{-2\pi\mu}$, thus we will henceforth neglect all contributions from NTO parts.
	
	\subsection{The left/right blob substructures}\label{npts}
	The above discussion is based on the early-time oscillation behavior of left/right blobs. Namely, the dynamical phase is linearly growing with the conformal time $\tau$. This will indeed be the case if the substructure of left/right blobs contains solely massless fields. Then the massive field will resonate with all possible frequencies contained in $F_{L,R}$. However, one can imagine that the substructure may also involve massive fields, which allow for dynamical phases growing logarithmically with conformal time. The resonance condition then becomes much more complicated than (\ref{resonbanceconstraint}), and new types of CC signals would appear. Physically, this corresponds to bulk processes of higher complexity, such as the cascade decay of massive particles. Nevertheless, these processes are heavily suppressed, and there are two ways to treat these extra massive propagators if we focus on the leading CC signal.
	
	First, if the massive propagator has nothing to do with the CC signal under consideration, there will be no resonance at the corresponding vertices. Thus it can be integrated out using a partial EFT. In other words, we mimic its effect by some local operators, whose contribution is perturbative in $\mu^{-1}$ and analytic in the momenta.
	
	Second, if the massive propagator does resonate at the corresponding vertices, the whole diagram will gain more suppression factors (each extra resonant massive propagator contributes an $e^{-\pi\mu}$ suppression). Thus, we can ignore this possibility at leading order safely.
	
	To summarize, our strategy is to work with one resonant massive propagator at a time, while contracting other massive propagators to local EFT operators and absorbing them into the left/right blobs. We then traverse every massive propagator in the tree diagram and sum their CC signals together. This should take care of all leading-order signals non-perturbative in $\mu^{-1}$. In the end, we can integrate out all massive fields and put back the total EFT tower as a background. 
	
	Now that we have reduced all massive modes in the left/right blobs, we can work out the possible forms of $k_L$ and $k_R$, and hence that of the CC signals. Although their detailed expressions will depend on the substructure of the left/right blobs shown in the Fig.(\ref{tree}), their general forms can be found via mathematical induction. We can prove that in the absence of IR divergences, the time dependence of the left/right blobs is always in the following form:
	\begin{equation}
		\mathcal{F}_{L}(\tau)=\sum_{k_L}\mathcal{P}_L(\tau)e^{i k_L\tau}\quad,\quad \mathcal{F}_{R}(\tau')=\sum_{k_R}\mathcal{P}_R(\tau')e^{i k_R\tau'}~,
	\end{equation}
	where $\mathcal{P}_{L,R}$ are polynomials. The summation on $k_{L,R}$ ranges over all possible frequencies contained in the left/right blob. The proof is quite straightforward.	First, notice the integral
	\begin{equation}
		\int x^n e^{ix}dx=(-i)^{3 n+1} \Gamma (n+1,-i x)~.
	\end{equation}
	For a positive integer $n\in \mathbb{N}$ (which is always the case for IR convergent massless fields), the incomplete gamma function can be simplified recursively to a polynomial multiplied by an exponential function,
	\begin{align}
		\nonumber\Gamma(n+1,-ix)&=n\Gamma(n,-ix)+(-ix)^n e^{ix}\\
		\nonumber&=\cdots\\
		\nonumber&=n!\Gamma (1,-ix)+\cdots+(-ix)^n e^{ix}\\
		&=\left[n!+\cdots+(-ix)^n\right] e^{ix}~.\label{PolynomialReduction}
	\end{align}
	Now, we focus on the substructure of the left blob.
	\begin{figure}[htp] 
		\centering
		\includegraphics[width=10cm]{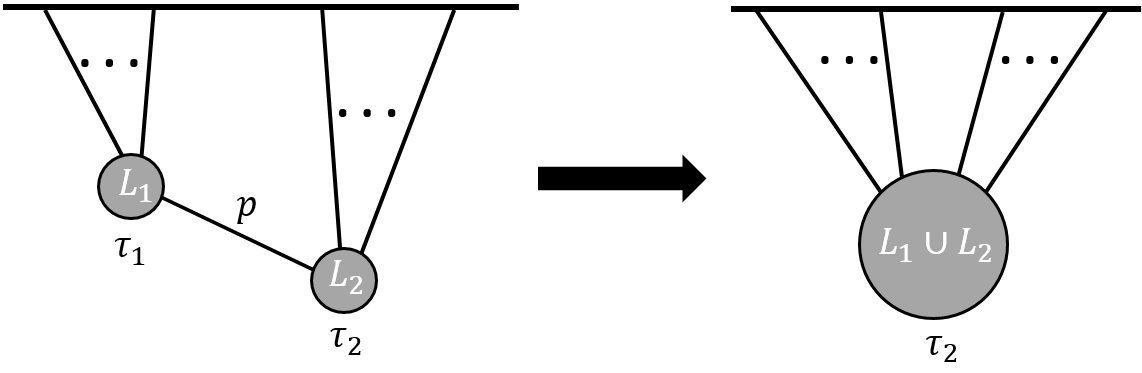}
		\caption{A larger left blob $L_1\cup L_2$ can be constructed by gluing the roots of two smaller blob trees with a massless propagator of momentum $p$.}\label{leftrightblob}
	\end{figure}
	We start with the assumption that, a left blob with $n_{L_1}$ external lines takes the form
	\begin{equation}
		\mathcal{F}_{L_1}(\tau_1)=\sum_{k_{L_1}}\mathcal{P}_{L_1}(\tau_1)e^{i k_{L_1}\tau_1}~,
	\end{equation}
	with $\mathcal{P}_{L_1}$ being a polynomial and $k_{L_1}$ traversing all frequencies in $L_1$. This is obviously true for the simplest blob substructure, namely an $n_{L_1}$-point contact vertex. More complex left blobs can be constructed by gluing two smaller left blobs. For instance, consider another left blob $L_2$, with
	\begin{equation}
		\mathcal{F}_{L_2}(\tau_2)=\sum_{k_{L_2}}\mathcal{P}_{L_2}(\tau_2)e^{i k_{L_2}\tau_2}~.
	\end{equation}
	Gluing the root of the two trees together by a massless propagator of internal momentum $\mathbf{p}=\sum_{i=1}^{n_{L_1}}\mathbf{k}_{i}$, we have
	\begin{align}
		\mathcal{F}_{L_1\cup L_2}(\tau_2)&=\int_{-\infty}^0 d\tau_1 \mathcal{F}_{L_1}(\tau_1)\hat{V}_1(\tau_1)G_{++}(\mathbf{p},\tau_1,\tau_2)\hat{V}_2(\tau_2)\mathcal{F}_{L_2}(\tau_2)~,
	\end{align}
	where $\hat{V}_{1,2}$ are interaction vertices that may contain integer multiple of derivative operators and scale factors. The spatial derivatives factor out from the integral, while time derivatives only change the detailed form of the polynomial, which is unimportant here. Hence, we can reduce the blob $L_1\cup L_2$ as follows,
	\begin{align}
		\nonumber\mathcal{F}_{L_1\cup L_2}(\tau_2)&= \sum_{k_{L_1},k_{L_2}}\Bigg[\int_{-\infty}^{\tau_2}d\tau_1 \hat{\mathcal{P}}_{L_1}(\tau_1)e^{i (k_{L_1}+p)\tau_1}\hat{\mathcal{P}}_{L_2}(\tau_2)e^{i (k_{L_2}-p)\tau_2}\\
		\nonumber&\quad\quad\quad\quad\quad+\int_{\tau_2}^{0}d\tau_1 \hat{\mathcal{P}}_{L_1}(\tau_1)e^{i (k_{L_1}-p)\tau_1}\hat{\mathcal{P}}_{L_2}(\tau_2)e^{i (k_{L_2}+p)\tau_2}\Bigg]\\
		\nonumber&= \sum_{k_{L_1},k_{L_2}} \Bigg[\mathcal{Q}(\tau_2)e^{i (k_{L_1}+p)\tau_2}\hat{\mathcal{P}}_{L_2}(\tau_2)e^{i (k_{L_2}-p)\tau_2}\\
		\nonumber&\quad\quad\quad\quad\quad+\mathcal{R}(\tau_2)e^{i (k_{L_1}-p)\tau_2}\hat{\mathcal{P}}_{L_2}(\tau_2)e^{i (k_{L_2}+p)\tau_2}+\mathcal{S}(\tau_2)\hat{\mathcal{P}}_{L_2}(\tau_2)e^{i (k_{L_2}+p)\tau_2}\Bigg]\\
		&=\sum_{k_{L_1},k_{L_2}}\Bigg[\left(\mathcal{Q}(\tau_2)+\mathcal{R}(\tau_2)\right)\hat{\mathcal{P}}_{L_2}(\tau_2)e^{i (k_{L_1}+k_{L_2})\tau_2}+\mathcal{S}(\tau_2)\hat{\mathcal{P}}_{L_2}(\tau_2)e^{i (p+k_{L_2})\tau_2}\Bigg]~.\label{mathinductionsteps}
	\end{align}
	Here $\hat{\mathcal{P}}_{L_1},\hat{\mathcal{P}}_{L_2},\mathcal{Q},\mathcal{R},\mathcal{S}$ are polynomials whose detailed forms are of no importance. Notice that in the second step of (\ref{mathinductionsteps}), we have formally performed the integral over $\tau_1$, using the fact that such integrals can be reduced as in (\ref{PolynomialReduction}), to polynomials multiplied by an exponential function. Interestingly, after integrating out $\tau_1$, two different exponential factors appear. The first one depends on the injecting frequency ($k_{L_1}+k_{L_2}$), the other one depends on the internal momentum ($p+k_{L_2}$). In this way, we have shown that gluing two smaller left blobs gives rise to a larger left blob that shares the same form of time dependence,
	\begin{equation}
		\mathcal{F}_{L_1\cup L_2}(\tau_2)=\sum_{k_{L_1\cup L_2}}\mathcal{P}_{L_1\cup L_2}(\tau_2)e^{i k_{L_1\cup L_2}\tau_2}~,
	\end{equation}
	where the summation ranges over $k_{L_1\cup L_2}=k_{L_1}+k_{L_2},p+k_{L_2}$ for all possible values of $k_{L_1},k_{L_2}$. 
	
	By iterating this gluing procedure many times, one can find all possible oscillation frequencies of the left blob $L=L_1\cup L_2\cup\cdots$. The right blob is treated the same way. Hence all CC signal patterns can be found out as
	\begin{eqnarray}
		&&\text{Non-Local Type} \quad \left[\frac{\mathcal{T}(\{k_L\}\mathcal{T}(\{k_R\}))}{\mathcal{T}(\{k_I\})}\right]^{\mp i\mu}~,\\
		&&\text{Local Type}\quad \left[\frac{\mathcal{T}(\{k_L\})}{\mathcal{T}(\{k_R\})}\right]^{\mp i\mu}~,
	\end{eqnarray}
	where $\mathcal{T}$ denotes all possible combinations of frequency sets.
	
	\subsection{Summary of the cutting algorithm}\label{CuttingRuleSummary}
	In this subsection, we summarize the procedures of applying our cutting rule for CC signals. For a $n$-point correlation function, denote the set of all massive propagators as $\Sigma$.
	
	\begin{framed}
		{\noindent {\it Cutting algorithm for CC signals}
			\begin{enumerate}
				\item Given a specific tree diagram, consider only single-colored diagrams with vertices being all-black ($+$) or all-white ($-$). Any mixed-colored diagrams are negligible by at least a factor $\mathcal{O}(e^{-2\pi\mu})$.
				\item Focus on one massive propagator $I\in\Sigma$ and integrate out all other massive propagators as local EFT operators. This reduces the tree topology to that of Fig.~\ref{tree}.\label{Step2}
				\item Explicitly compute the left/right blobs. The results should be reducible to $\mathcal{F}_L(\tau)=\sum_{k_L}\mathcal{P}_L(\tau)e^{ik_L\tau}$ and $\mathcal{F}_R(\tau')=\sum_{k_R}\mathcal{P}_R(\tau')e^{ik_R\tau'}$ in the absence of IR divergences, where $\mathcal{P}_{L,R}$ are polynomials, and the sums over $k_{L,R}$ take into account of all possible frequencies.
				\item For $k_L>k_R$, flip $\theta(\tau-\tau')=1-\theta(\tau'-\tau)$ and perform two factorized integrals,
				\begin{equation}
					S_{I}^>(k_L,k_R)=\int_{-\infty}^{0}d\tau \mathcal{F}_L(\tau)v_{k_I}(\tau)\times\int_{-\infty}^{0}d\tau' \mathcal{F}_R(\tau')v_{k_I}^*(\tau')~.
				\end{equation}
				The integral takes the form of a Laplace transformation of special functions, which often yields analytical expressions,
				\begin{equation}
					S_{I}^>(k_L,k_R)=\mathcal{L}\{\mathcal{P}_L v_{k_I}\}(i k_I)\times\mathcal{L}\{\mathcal{P}_R v_{k_I}^*\}(i k_I)~.
				\end{equation}
				The leftover commutator integral has no CC signals in it, and can be discarded.
				\item Account for the case $k_L<k_R$ by symmetrization, and traverse all possible frequencies,
				\begin{equation}
					S_I=\sum_{k_L,k_R}\left[\theta(k_L-k_R)S_{I}^>(k_L,k_R)+\theta(k_R-k_L)S_{I}^>(k_R,k_L)\right]~.
				\end{equation}
				\item Repeat Step 2-5 for each massive propagator $I\in \Sigma$, then sum them up to obtain the total signal,
				\begin{equation}
					S=\sum_{I\in\Sigma} S_I~.
				\end{equation}
				\item At last, dress the total signal $S$ by the EFT background $B$ where all massive fields are integrated out.\label{Step7}
				\begin{equation}
					\langle\zeta^n\rangle\simeq (S+B)\sim \Big[\mathcal{O}(|\beta|)\times (\text{non-local})+\mathcal{O}(e^{-\pi\mu})\times (\text{local})\Big]+\mathcal{O}\left(\frac{\text{EFT}}{\mu^{\#}}\right)~.
				\end{equation}
			\end{enumerate}
		}
	\end{framed}
	Before going to explicit examples, we make a few comments on this cutting algorithm.
	\begin{enumerate}
		\item The cut result $S+B$ is only an approximation to the full $n$-point function. The error comes from two sources. First, to obtain closed expressions, one must truncate the large-mass EFT in both Step~\ref{Step2} and Step~\ref{Step7}. This results in errors suppressed by powers of $\mu^{-2}$, which can be reduced if we include more EFT operators. Second, we have thrown away all the commutator integrals since they do not contain CC signals. However, they are important for restoring differentiability at $k_L=k_R$ and canceling the folded limit pole at $k_{L,R}\to k_I$. This results in an $\mathcal{O}(e^{-\pi\mu})$ error in these special limits, one that cannot be reduced within our method. Yet in practice, as long as we stay away from these special limits (which only take up a small fraction of the total phase space), this source of error is generically small, as will be demonstrated in Sect.~\ref{ApplicationSect}.
		
		\item The Laplace transformation of familiar mode functions (Hankel, Whittaker) can be evaluated to hypergeometric functions. This would not have been possible without cutting the nested time integral. In addition, EFT contributions are also analytically calculable. Thus our cutting rule typically results in completely analytical results\footnote{Yet, necessarily non-analytic.}. Even if the Laplace transform of unfamiliar mode functions may not be evaluated in a closed form, the original two-dimensional integral is now factorized into two one-dimensional integrals. This greatly reduces the computational cost of numerical integration.
		
		\item Each step in the above algorithm is specific but tedious to perform by hand. Therefore, the best way to systematically carry out the cutting procedure is to promote this into a computer program, a goal that we hope to accomplish in the future. This would be beneficial for model builders in cosmological collider physics. 
	\end{enumerate}
	Finally, a comparison must be made between our method and the recently proposed cosmological cutting rules \cite{Goodhew:2020hob,Melville:2021lst,Goodhew:2021oqg,Baumann:2021fxj,Meltzer:2021zin}. Based on general principles such as locality, unitarity and analyticity, the cosmological cutting rules give precise relations among the discontinuities of wavefunction exponents $\text{Disc}_{k_I}\psi_n$. From these relations, one may recursively construct the whole wavefunction. In contrast, our cutting rule directly applies to correlation functions $\langle\zeta^n\rangle$. It is based on the physical picture of bulk evolution with the specific aim of extracting CC signals of massive fields. Our cut result is also approximate, with reducible EFT truncation errors and irreducible errors from neglecting commutator integrals. However, as stated above, the algorithm itself is well-defined and can be readily implemented as computer programs yielding analytical expressions. We have traded mathematical rigor and formality for practicality and efficiency.
	
	Of course, there are also similarities. The fact that the non-local CC signal is associated with a branch cut discontinuity in $k_I$ and that it actually comes from a factorized integral seem to hint a connection to the cosmological cutting rules. 
	It is certainly interesting to further explore the similarities and dissimilarities of these two approaches in the future. 
	
	\section{Application to Typical Diagrams}\label{ApplicationSect}
	In this section, we will explicitly demonstrate how to apply our cutting rule to extract CC signals for the three typical diagrams shown in Fig.~\ref{typicaldiam}. For a better illustration, all diagrams calculated in this section correspond to the curvature trispectrum (4-point function). The oft studied bispectrum (3-point function) can be easily obtained by taking the soft limit of one external momentum, and we give more related discussions at the end of Sect.~\ref{sec41}.
	\subsection{Example 1: Direct scalar exchange}\label{sec41}
\begin{figure}[htp] 
		\centering
		\includegraphics[width=8cm]{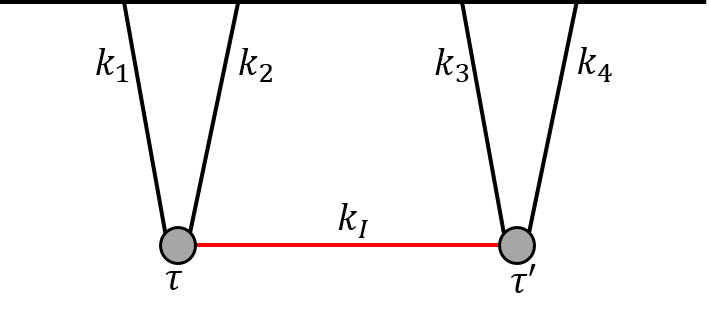}
		\caption{\label{example1} Example 1: A direct massive scalar exchange.}
	\end{figure}

	The first example (Fig.~\ref{example1}) is that of a simple exchange diagram with an intermediate massive scalar $\sigma$ of mass $\frac{m}{H}=\sqrt{\mu^2+\frac{9}{4}}$. The interaction vertex is chosen to be
	\begin{equation}\label{inta}
	\mathcal{L}_3= c_3 a^2 \varphi'^2\delta\sigma~,
	\end{equation}
	where $\varphi$ is the massless inflaton.
	
	Setp 1, according to the cutting algorithm, we only focus on the TO diagram with the vertices being black. The ATO diagram is obtained via complex conjugation and momenta reversal. More explicitly,
	\begin{align}
	\nonumber{\langle\zeta_\mathbf{k_1}\zeta_\mathbf{k_2}\zeta_\mathbf{k_3}\zeta_\mathbf{k_4}\rangle}'_{s,\text{TO}}=& -\tilde{c_3}^2\int_{-\infty}^{0}\int_{-\infty}^{0}d\tau d\tau'a^2(\tau)a^2(\tau')\partial_{\tau}G_{++}(\mathbf{k_1},\tau,0)\partial_{\tau}G_{++}(\mathbf{k_2},\tau,0)\\ &\times D_{++}(\mathbf{k}_I,\tau,\tau')\partial_{\tau'}G_{++}(\mathbf{k_3},\tau',0)\partial_{\tau'}G_{++}(\mathbf{k_4},\tau',0)~.
	\end{align}
	Here $\mathbf{k}_I=\mathbf{k}_1+\mathbf{k}_2$, $\tilde{c_3}\equiv\frac{ c_3\dot{\phi_0}^2}{H^2}$ and $G,D$ are the propagators for $\varphi,\sigma$, respectively. Without loss of generality, we take the $s$-channel as an example. The $t,u$-channels can be included after permutation.
	
	Step 2 and 3 are trivial since no other massive propagators exist in this simple diagram. The left/right blobs are
	\begin{eqnarray}
		\mathcal{F}_L(\tau)&=&2i\tilde{c_3}u_{k_1} u_{k_2}(0)a^2u'^*_{k_1}u'^*_{k_2}(\tau)~,\\
		\mathcal{F}_R(\tau')&=&2i\tilde{c_3}u_{k_3} u_{k_4}(0)a^2u'^*_{k_3}u'^*_{k_4}(\tau')~.
	\end{eqnarray}
	
	Then at Step 4, denoting $k_{i_1\cdots i_m}\equiv k_{i_1}+\cdots+k_{i_m}$, we take $k_{12}>k_{34}$ and flip one of the Heaviside step functions to split the TO diagram into
	\begin{eqnarray}
	{\langle\zeta^4\rangle}_{s,TO}&=&S^>_{I}+\mathcal{I}_{I,com}~,
	\end{eqnarray}
	with the cut result
	\begin{align}\label{triccs}
	\nonumber S^>_{I}(k_{12},k_{34})&=\int_{-\infty}^{0}d\tau\mathcal{F}_L v_{k_I}(\tau)\int_{-\infty}^{0}d\tau'\mathcal{F}_R v^*_{k_I}(\tau')\\
	&=-4\tilde{c_3}^2u_{k_1} u_{k_2}u_{k_3}u_{k_4}(0)\left[\int_{-\infty}^{0}d\tau a^2u'^*_{k_1}u'^*_{k_2}v_{k_I}\int_{-\infty}^{0}d\tau' a^2u'^*_{k_3}u'^*_{k_4}v^*_{k_I}\right]~,
	\end{align}
	and the commutator integral
	\begin{align}\label{tricom}
	\nonumber \mathcal{I}_{I,com}=&\int_{-\infty}^{0}d\tau\int_{\tau}^{0}d\tau'\mathcal{F}_L(\tau) \mathcal{F}_R(\tau') \left(v^*_{k_I}(\tau)v_{k_I}(\tau')-v_{k_I}(\tau)v^*_{k_I}(\tau')\right)\\
	=&-\tilde{4 c_3}^2u_{k_1} u_{k_2}u_{k_3}u_{k_4}(0)\nonumber\\
	&\times\int_{-\infty}^{0}d\tau a^2 u'^*_{k_1}u'^*_{k_2}(\tau)\int_{\tau}^{0}d\tau'a^2 u'^*_{k_3}u'^*_{k_4}(\tau')\left(v^*_{k_I}(\tau)v_{k_I}(\tau')-v_{k_I}(\tau)v^*_{k_I}(\tau')\right)~.
	\end{align}
	The CC signals can be evaluated via Laplace transformation as\footnote{The negligible NTO diagram also have a similar form, more details can be found in Appendix~\ref{intcc}.}
	
	\begin{align}
		\nonumber S^>_{I}(k_{12},k_{34})&=-\left(\frac{H}{\dot{\phi_0}}\right)^4\frac{\pi^2{c_3}^2H^6(16\mu^4+40\mu^2+9)^2\sech^2(\pi\mu)}{2^{17}k_1k_2k_3k_4k_I^5}\\
		&\times{}_2F_1\Bigg[\begin{array}{c} \frac{5}{2}-i\mu, \frac{5}{2}+i\mu\\[2pt] 3 \end{array}\Bigg|\,\frac{k_I+k_{12}}{2k_I}\Bigg]\times {}_2F_1\Bigg[\begin{array}{c} \frac{5}{2}-i\mu, \frac{5}{2}+i\mu\\[2pt] 3 \end{array}\Bigg|\,\frac{k_I-k_{34}}{2k_I}\Bigg]~.
	\end{align}
   Here $k_{12}$ is assumed to have a small negative imaginary part in order to select the correct branch of the hypergeometric function. As argued above, the commutator integral contains neither local nor non-local signals. We numerically verify this fact in Appendix~\ref{prcmint}.
	
	Step 5, the total cut CC signals is obtained via symmetrization,
	\begin{equation}
		S_I=\theta(k_{12}-k_{34})S_{I}^>(k_{12},k_{34})+\theta(k_{34}-k_{12})S_{I}^>(k_{34},k_{12})+\text{c.c.}~.\label{firstdiagsymtrS}
	\end{equation}

	Step 6 is trivial and Step 7 is to supplement the CC signals with an EFT background. This background admits a perturbative expansion in powers of $\mu^{-1}$. These are analytic functions of momenta that can be mimicked by local operators in the single-field EFT. The EFT background is usually smaller than CC signals under soft limits. Nevertheless, it plays an important role in ``equilateral'' regions where all momenta (both external and internal) are comparable in magnitude. In this example, the leading order EFT contribution is generated from the operator
	\begin{align}
	\Delta\mathcal{L}^{EFT}_I=\frac{c_3^2}{2\mu^2 H^2}\varphi'^4~.
	\end{align}
	In the ``$s$''-channel, its impact on the 4-point function is
	\begin{align}\label{eft}
	B_I=\left(\frac{H}{\dot{\phi_0}}\right)^4\frac{12H^6c_3^2}{\mu^2k_1k_2k_3k_4k_{1234}^5}~.
	\end{align}
    Finally, we get the analytical approximation for the 4-point function in the $s$-channel:
	\begin{align}
	\langle\zeta_\mathbf{k_1}\zeta_\mathbf{k_2}\zeta_\mathbf{k_3}\zeta_\mathbf{k_4}\rangle_s'\approx S_{I}+B_I~,
	\end{align}
	with $S_I$ and $B_I$ given by (\ref{firstdiagsymtrS}) and (\ref{eft}).
	
	\begin{figure}[htp]
		\centering
		\includegraphics[width=14.8cm]{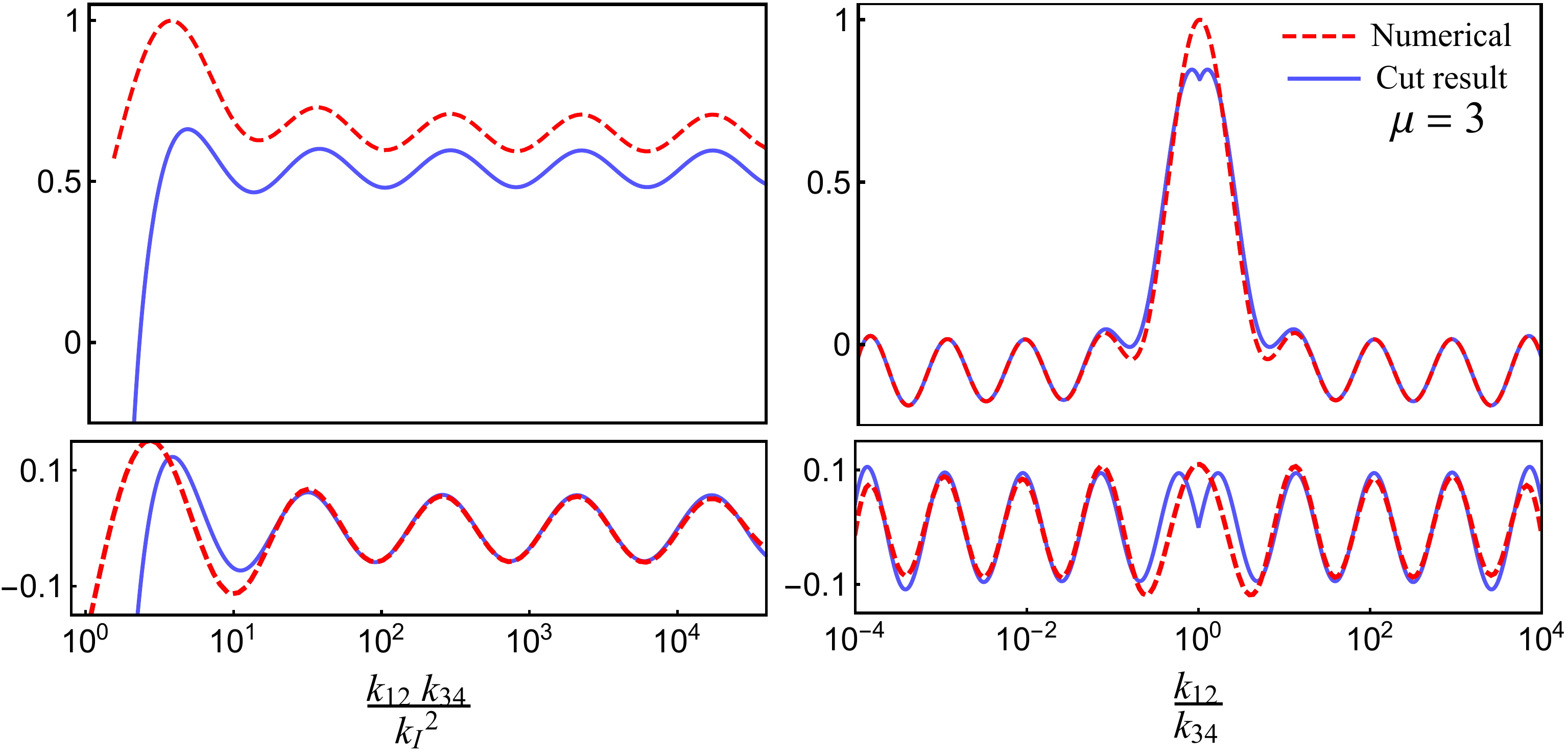}
		\caption{The comparison between the 4-point functions obtained from the cutting rule (blue lines) and numerical integration (red dashed lines). \textit{Left panel}: Non-local type CC signals, where we have fixed $k_{12}/k_{34}=1$, and multiplied a factor $\left(k_{12}k_{34}/k_I^2\right)^{9/2}$ for better visualization. \textit{Right panel}: Local type CC signals, where we have fixed $k_{12}k_{34}/k_I^2=4\times10^4$. All results in the upper panels are normalized by their maxima. In the lower panels, we compare the CC signals filtered from the numerical result and those directly obtained via the cut. We have adopted a high-pass filter with Hamming window.}\label{eft+cc}
	\end{figure}
	In Fig.~\ref{eft+cc}, we compare our cut result to that of brute-force numerical integration. The left column shows the non-local type CC signal where we keep $k_{12}\approx k_{34}$ and vary $\frac{k_{12}k_{34}}{k_I^2}$. The right column shows the local type signal where we keep $\frac{k_{12}k_{34}}{k_I^2}=4\times 10^4$ and vary $\frac{k_{12}}{k_{34}}$. The upper panels show the full 4-point function while the lower panels show the CC signals alone. In order to extract oscillatory signals from the total numerical result, we have evoked \textit{high-pass filters} to eliminate the low-frequency EFT background \cite{Wang:2021qez}. As we can see from the plots, the numerical results (red dashed lines) match our analytical cut results (blue line) very well, especially for the signal part. The slight mismatch comes around in two places. First, when $k_{12}\approx k_{34}$, the overall signal size differ relatively by $\mathcal{O}(\mu^{-2})\sim\frac{1}{9}$, which is the next-order contribution in the EFT tower. We expect that including more EFT operators induced by the exchange diagram can systematically cure this mismatch. The second type of mismatch manifest itself in the folded limit ($k_{12},k_{34}\to k_I$) pole as well as in the derivative discontinuity at $k_{12}=k_{34}$. As mentioned before, these are caused by the disposal of the commutator integral, which, despite being small, plays an important role in restoring BD vacuum in the early-time limit. This intrinsic mismatch shows a defect of our method. Nevertheless, as long as we stay away from the special configurations, the CC signals we care about will not be influenced much.
	
	Finally, we can gain some insight by studying the large-mass asymptotic behavior of $S_I$ in the soft limit,
	\begin{equation}
	S_I^>\xrightarrow[k_I\to 0]{\mu\gg 1}~\propto \mu ^3 e^{- \pi\mu } \left[\cos \left(\mu  \ln \frac{{4k_{12}k_{34}}}{k_I^2}\right)-\sin \left(\mu  \ln
	\frac{k_{12}}{k_{34}}\right)\right]~,\label{cto}
	\end{equation}
	where we have only kept the leading contribution. Clearly, there are two independent oscillation patterns with degenerate amplitudes. In the literature, this large-mass result is often obtained via direct late-time (IR) expansion (\ref{MassivePropIR}). Here, we see that the late-time expansion is qualitatively justified for large masses in this 4-point function case. However, there are two disadvantages for the traditional late-time expansion. First, while the late-time expansion gives the correct exponential suppression factor $e^{-\pi\mu}$, it may fail to capture the correct mass power dependence ($\mu^n$), which is also important in the estimation of signal strength. Such is the case, for example, in the 3-point diagram frequently encountered in cosmological collider physics. If one performs a late-time expansion before time integration \cite{Chua:2018dqh}, the CC signal amplitude is proportional to $\mu^2 e^{-\pi\mu}$. However, as we can see from tending $k_4\to 0$ in (\ref{cto}), the correct amplitude should be $\mu^3 e^{-\pi\mu}$. This can reproduce the results of~\cite{Chen:2015lza}. The problem arises because performing a late-time expansion does not commute with completing the time integral. This fact was previously pointed out in \cite{Wang:2020ioa}. Second, the signal waveform obtained from the late-time expansion is only valid in the soft limit $k_I\to 0$. Significant dephasing of the waveform will appear when the internal momentum is less soft. In other words, its validity is limited to a small part of the phase space. In contrast to the late-time expansion, our method can get the correct amplitude as well as the correct waveform of the CC signals, greatly enlarging the phase space we can take advantage of.

	\subsection{Example 2: Scalar exchange with two-point mixing and different masses}
\begin{figure}[htp] 
		\centering
		\includegraphics[width=8cm]{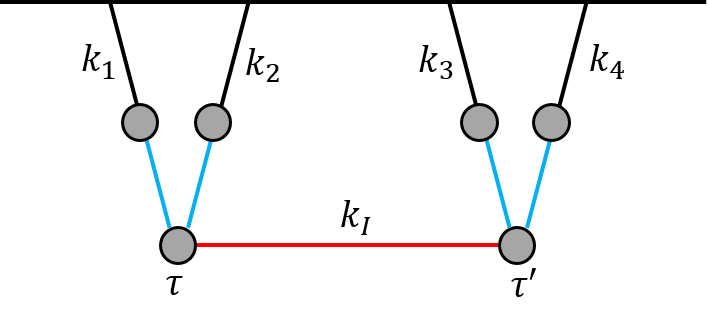}
		\caption{\label{example2} Example 2: Massive scalar exchange with two-point mixing.}
	\end{figure}

	The second diagram (Fig.~\ref{example2}) involving two-point mixing is considerably more complicated. As described in Sect.~\ref{CCPReview}, the TO diagram alone consists of a 6-layer integral. However, with the cutting rule, we can extract the CC signals without too much effort, as we will demonstrate below.
	
	Let the mass of exchanged massive field $\sigma$ (red line) be $\mu$ and the mass of the massive field $s$ on the external lines (cyan) be $\nu$. The interactions are set by the following couplings,
	\begin{align}
		\Delta\mathcal{L}_2=\rho a^3 \varphi's~~,~~\Delta\mathcal{L}_3=\lambda a^4 s^2\sigma~.
	\end{align}
	Without loss of generality, we assume that all propagators are made distinguishable. It is not hard to include permutations according to the Bose statistics.
	
	First, focusing on the TO diagram with all vertices being black, the correlation function reads
	\begin{align}
		&\nonumber{\langle\zeta_\mathbf{k_1}\zeta_\mathbf{k_2}\zeta_\mathbf{k_3}\zeta_\mathbf{k_4}\rangle}'_{s,\text{TO}}\\
		=&-\lambda^2\left(\frac{H}{\dot{\phi_0}}\right)^4\int_{-\infty}^{0}d\tau\tau'a^4(\tau)a^4(\tau')\mathcal{G}_+(k_1,\tau)\mathcal{G}_+(k_2,\tau)\mathcal{G}_+(k_3,\tau')\mathcal{G}_+(k_4,\tau')D^\sigma_{++}(k_I,\tau,\tau')~.
	\end{align}
	The effective propagator $\mathcal{G}$ is defined as \cite{Chen:2018sce}
	\begin{align}
		\mathcal{G}_{\pm}(k,\tau_1)=i\rho\int_{-\infty}^{0}d\tau_2a^3(\tau_2)\left[D^s_{++}(k,\tau_1,\tau_2)\partial_{\tau_2}G_{++}(k,\tau_2,0)\right]~,
	\end{align}
	where $D^s$ and $D^\sigma$ are the massive propagators of the field $s$ and the field $\sigma$. 
	
	\begin{figure}[htp]
		\centering
		\includegraphics[width=14cm]{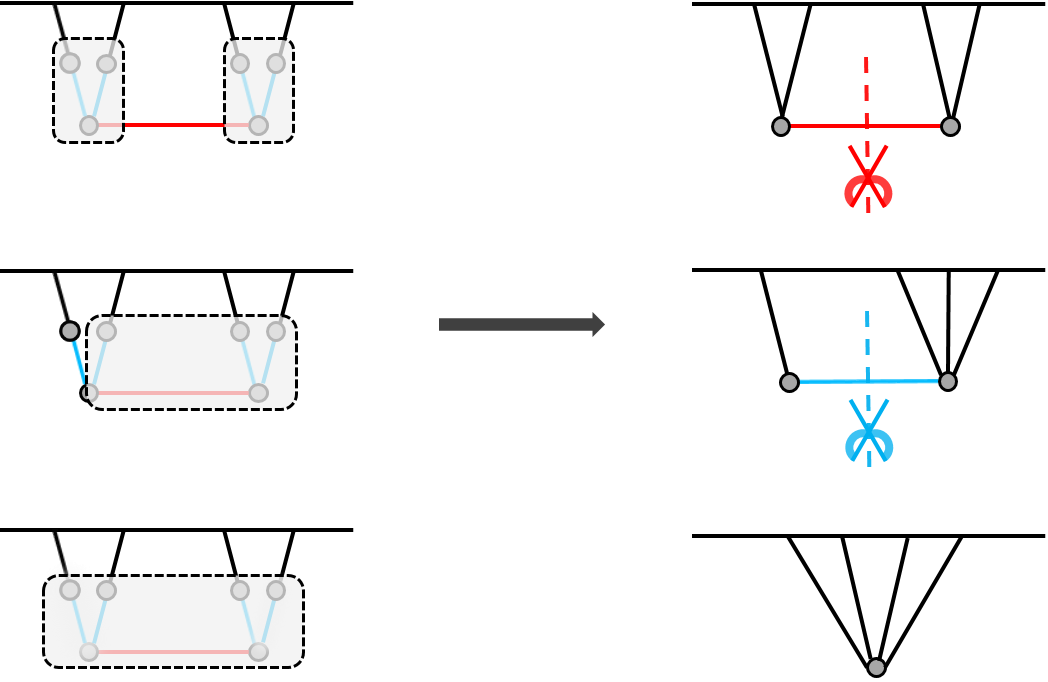}
		\caption{The strategy of the cutting algorithm for the diagram in Fig.~\ref{example2}. We traverse all massive propagators and cut one at a time ($(12,34)$ on the first row and $(1,234)$ on the second row), while integrating out others as EFTs. At last, we integrate all of them to obtain a total EFT background $B$.}\label{partialeft}
	\end{figure}
	Step 2, there are five massive propagators in this diagram. Naming them according to the external lines separated by each one of them, we have the set of all massive propagators,  
	\begin{equation}
		\Sigma=\Big\{(12,34),~(1,234),~(2,134),~(3,124),~(4,123)\Big\}~.
	\end{equation}
	We focus on one massive propagator $I\in\Sigma$ at a time, integrating out the others as EFT operators. This whole procedure is illustrated in Fig.~\ref{partialeft}. For example, doing so for $I=(12,34)$ yields a contact EFT vertex at leading order,
	\begin{align}
		\Delta\mathcal{L}^{EFT}_{(12,34)}=\frac{\lambda\rho^2}{(H\nu)^4}a^2\varphi'\varphi'\sigma~.
	\end{align}
	Step 3, the left/right blobs are now easily computable:
	\begin{eqnarray}
		\mathcal{F}_L(\tau)&=&\frac{i\lambda\rho^2}{(H\nu)^4}u_{k_1} u_{k_2}(0)a^2u'^*_{k_1}u'^*_{k_2}(\tau)~,\\
		\mathcal{F}_R(\tau')&=&\frac{i\lambda\rho^2}{(H\nu)^4}u_{k_3} u_{k_4}(0)a^2u'^*_{k_3}u'^*_{k_4}(\tau')~.
	\end{eqnarray}
	Step 4, for $k_{12}>k_{34}$, we cut the massive propagator $I=(12,34)$ by
	\begin{eqnarray}
		{\langle\zeta^4\rangle}_{(12,34),\rm{TO}}&=&S^>_{(12,34)}+\mathcal{I}_{(12,34),com}~,
	\end{eqnarray}
	with the cut result
	\begin{align}
		\nonumber S^>_{(12,34)}(k_{12},k_{34})&=\left(\frac{H}{\dot{\phi_0}}\right)^4\int_{-\infty}^{0}d\tau\mathcal{F}_L v_{k_I}(\tau)\int_{-\infty}^{0}d\tau'\mathcal{F}_R v^*_{k_I}(\tau')\\
		&=-\left(\frac{H}{\dot{\phi_0}}\right)^4\frac{\lambda^2\rho^4}{(H\nu)^8}u_{k_1} u_{k_2}u_{k_3}u_{k_4}(0)\left[\int_{-\infty}^{0}d\tau a^2u'^*_{k_1}u'^*_{k_2}v_{k_I}\int_{-\infty}^{0}d\tau' a^2u'^*_{k_3}u'^*_{k_4}v^*_{k_I}\right]~.
	\end{align}
	Step 5, symmetrization gives the complete CC signal in the channel $I=(12,34)$,
	\begin{equation}
		S_{(12,34)}=\theta(k_{12}-k_{34})S_{(12,34)}^>(k_{12},k_{34})+\theta(k_{34}-k_{12})S_{(12,34)}^>(k_{34},k_{12})+\text{c.c.}~.
	\end{equation}
	Step 6, we need to loop back to Step 2 and go through the same procedure for the other massive propagators $I\in\Sigma$. For instance, in the channel $I=(1,234)$, large-mass EFT expansion of other propagators gives the following local operator at leading order:
	\begin{align}
		\Delta\mathcal{L}^{EFT}_{(1,234)}=\frac{\lambda^2\rho^3}{H^8\mu^2\nu^6}a\varphi'\varphi'\varphi's~.
	\end{align}
	Thus the left/right blobs are
	\begin{eqnarray}
		\mathcal{F}_L(\tau)&=&i\rho u_{k_1}(0)a^3u'^*_{k_1}(\tau)~,\\
		\mathcal{F}_R(\tau')&=&\frac{i\lambda^2\rho^3}{H^8\mu^2\nu^6}u_{k_2} u_{k_3} u_{k_4}(0)au'^*_{k_2}u'^*_{k_3}u'^*_{k_4}(\tau')~.
	\end{eqnarray}
	After cut and symmetrization (which is trivial since $k_1<k_{234}$ by the triangle inequality), the corresponding CC signal in the channel $I=(12,34)$ is then
	\begin{equation}
		S_{(1,234)}=S_{(1,234)}^>(k_{234},k_{1})+\text{c.c.}~,
	\end{equation}
	with
	\begin{align}
		\nonumber S^>_{(1,234)}(k_{234},k_{1})&=\left(\frac{H}{\dot{\phi_0}}\right)^4\int_{-\infty}^{0}d\tau\mathcal{F}_L v_{k_I}^*(\tau)\int_{-\infty}^{0}d\tau'\mathcal{F}_R v_{k_I}(\tau')\\
		&=-\left(\frac{H}{\dot{\phi_0}}\right)^4\frac{\lambda^2\rho^4}{H^8\mu^2\nu^6}u_{k_1} u_{k_2}u_{k_3}u_{k_4}(0)\left[\int_{-\infty}^{0}d\tau a^3u'^*_{k_1}v^*_{k_I}\int_{-\infty}^{0}d\tau' a u'^*_{k_2}u'^*_{k_3}u'^*_{k_4}v_{k_I}\right]~.
	\end{align}
	Signals in the other channels differ from $(1,234)$ only by a rename of the external momenta. We need to sum all their contributions,
	\begin{equation}
		S=S_{(12,34)}+S_{(1,234)}+S_{(2,134)}+S_{(3,124)}+S_{(4,123)}~.
	\end{equation}
	Finally, in Step 7, we dress the signals with an overall EFT background. At leading order, it is sourced by the operator
	\begin{equation}
		\Delta\mathcal{L}^{EFT}_{\Sigma}=\frac{\lambda^2\rho^4}{H^{10}\mu^2\nu^8}\varphi'\varphi'\varphi'\varphi'~,
	\end{equation}
	which gives
	\begin{equation}
		B=\left(\frac{H}{\dot{\phi_0}}\right)^4\frac{3\lambda^2\rho^4 }{\mu ^2 \nu ^8H^{2}k_1 k_2 k_3 k_4 k_{1234}^5 }~.
	\end{equation}
	As a result, the cutting algorithm gives an analytical approximation to $\langle\zeta^4\rangle'$ as $\langle\zeta^4\rangle'=S+B$, with the detailed expressions of each $S_I$ shown in Appendix~\ref{AppExample2CCSignalExpression}.
	
	\begin{figure}[htp]
		\centering
		\includegraphics[width=15cm]{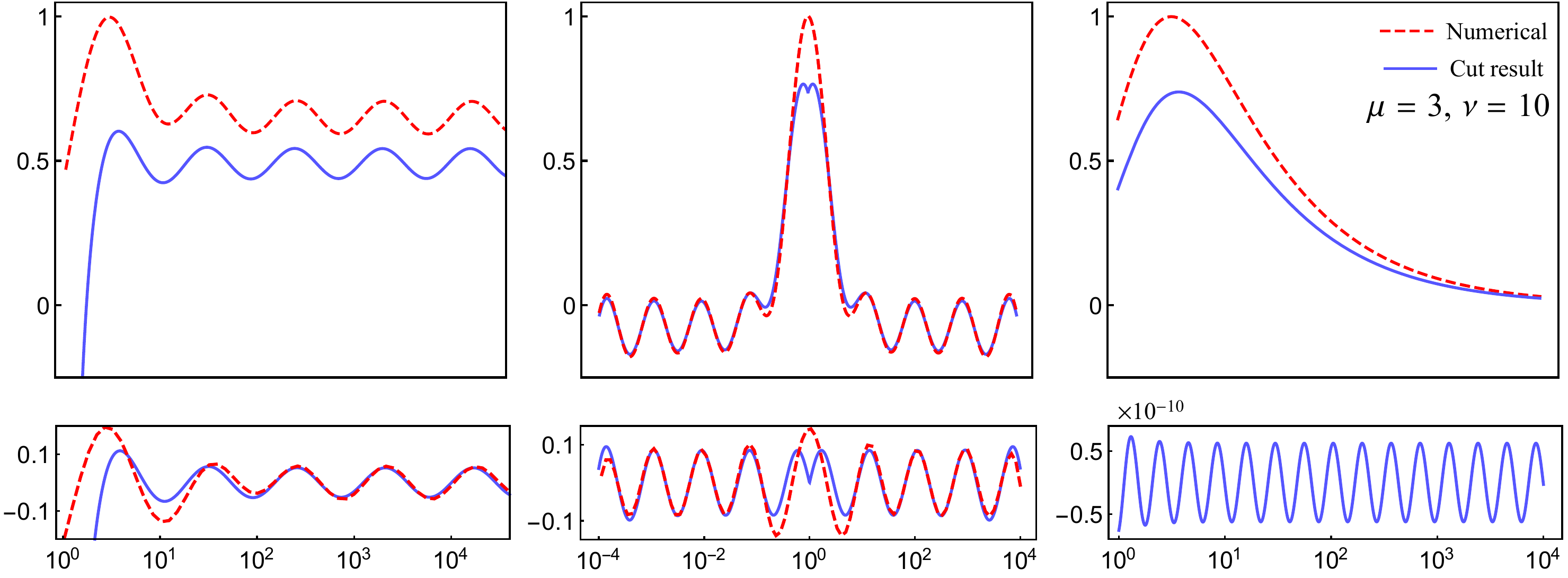}\\~\\
		\includegraphics[width=15cm]{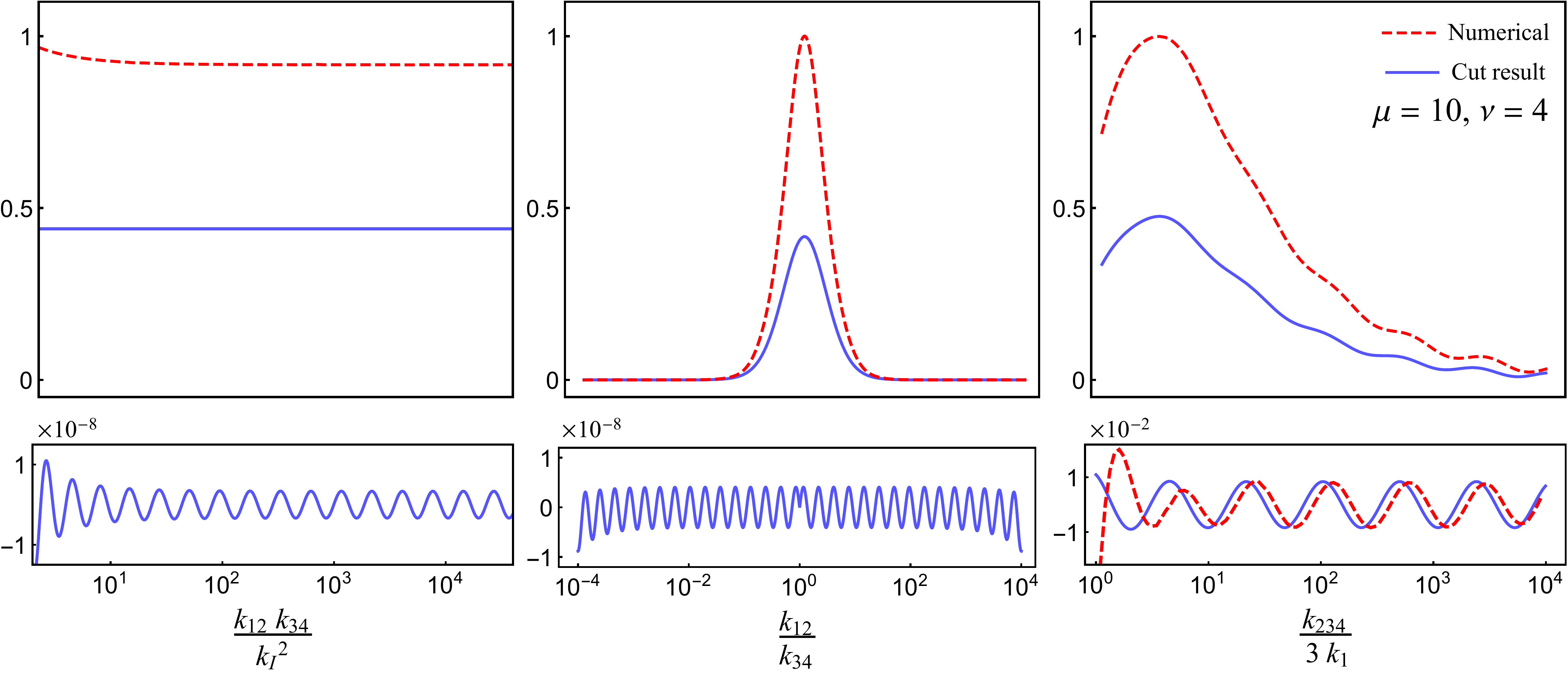}
		\caption{The 4-point functions obtained from the cut rule (blue lines) and numerical integration (red dashed lines). \textit{Left column}: Non-local type CC signals in the $(12,34)$-channel, where we have fixed $k_{12}/k_{34}=1$, and multiplied a factor $\left(k_{12}k_{34}/k_I^2\right)^{9/2}$ for better visualization. \textit{Middle column}: Local type CC signals in the $(12,34)$-channel, where we have fixed $k_{12}k_{34}/k_I^2=4\times10^4$. \textit{Right column}: Mixed type (non-local+local) CC signals in the $(1,234)$-channel, where we have fixed $k_2=k_3=k_4$. The first row corresponds to the mass choice $\mu=3,\nu=10$, while the second row corresponds to the mass choice $\mu=10,\nu=4$. The 4-point functions are normalized by their maxima. In the lower panels, we compare the CC signals filtered from the numerical result and that directly obtained via the cut. In cases where the CC signal strength is too weak, we have plotted the analytical cut result only, because the numerical accuracy is inadequate for signal filtering.}\label{seconddiagramComparison}
	\end{figure}
	
	We compare our cut result with numerics in Fig.~\ref{seconddiagramComparison}. We have plotted three types of CC signals for two sets of mass parameters. The upper panels correspond to the full 4-point function, while the lower panels show CC signals obtained via the cutting algorithm compared to that filtered from the numeric result. One can see that the oscillatory CC signals match well to that filtered from numerical integration, except at aforementioned special limits where our approximation breaks down. Yet the overall background has an $\mathcal{O}(1)$ mismatch. This is because in Step 2 and 7, we have only considered lowest order EFT operators induced by the massive fields\footnote{In fact, the EFT error receives amplification from the number of massive propagators. One can estimate the EFT error by $R\sim 2\times\left(\frac{1}{\mu^2}+\frac{4}{\nu^2}\right)$, which gives $R\sim 30\%$ and $R\sim 52\%$ for the two parameter choices in Fig.~\ref{seconddiagramComparison}. This is why it seems to be larger than that in Fig.~\ref{eft+cc}.}. This EFT error is also responsible for the small phase mismatch in the $\frac{k_{234}}{3k_{1}}$ signal. However, we expect such error can be systematically reduced by including higher-order EFT operators. Despite being tedious, the process of including more EFT operators is well-understood and can be performed in an analytical and mechanical fashion. We leave the methodical analysis of the EFT background to future studies.
	

	\subsection{Example 3: Vector exchange with chemical potential}\label{ChemPtlSect}
\begin{figure}[htp] 
		\centering
		\includegraphics[width=8cm]{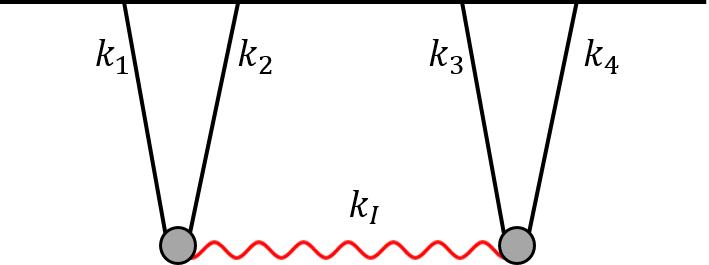}
		\caption{\label{example3} Example 3: Massive vector exchange with chemical potential.}
	\end{figure}

	The third example (Fig.~\ref{example3}) entails a spin-1 massive vector boson with chemical potential. The chemical potential of a vector boson $A_\mu$ can be introduced via coupling the gradient of a rolling scalar field $\theta(\tau)$ to the Chern-Simons current,
	\begin{equation}
		\Delta \mathcal{L}_{\text{chem}}=\frac{1}{2}\sqrt{-g}\kappa_\mu J_{CS}^\mu=-\frac{1}{4}\theta(\tau)\epsilon^{\mu\nu\rho\sigma}F_{\mu\nu} F_{\rho\sigma}+\text{total derivatives}~,
	\end{equation}
	where $J_{CS}^\mu=\frac{1}{\sqrt{-g}}\epsilon^{\mu\nu\rho\sigma}A_\nu F_{\rho\sigma}$ and $\kappa_\mu=\partial_\mu\theta(\tau)\equiv a(\tau) \kappa \delta_\mu^0$. The free-theory dynamics of the vector boson is then described by the Lagrangian
	\begin{align}
	\mathcal{L}_V=\sqrt{-g}\left(-\frac{1}{4}F_{\mu\nu}F^{\mu\nu}+\frac{1}{2}m^2A_\mu A^{\mu}\right)+\Delta\mathcal{L}_{\text{chem}}~.
	\end{align}
	The mode functions with helicity $\lambda=\pm,0$ are expressed in terms of the Whittaker function,
	\begin{align}
	 &v_{k_I}^{\pm}=\frac{1}{\sqrt{2k_I}}e^{\mp\pi\tilde\kappa/2}W_{\pm i \tilde\kappa,i\mu}(2ik_I\tau)~,\nonumber\\
	 &v_{k_I}^{0}=\frac{1}{\sqrt{2k_I}}\frac{1}{\sqrt{1+4\mu^2}}\left[2k\tau W_{0,1+i\mu}(2i k\tau )-(2\mu+i)W_{0,i\mu}(2ik\tau )\right]~,
	\end{align}
	where $\mu=\sqrt{\frac{m^2}{H^2}-\frac{1}{4}},\tilde{\kappa}\equiv\frac{\kappa}{H}$. The particle production amount can be computed using the Stokes-line method,
	\begin{equation}
		\langle n^\lambda_{\mathbf{k}_I}\rangle'|_{\tau\to 0}=|\beta^\lambda_{\mathbf{k}_I}(0)|^2\simeq e^{-\pi(\mu+\lambda\tilde\kappa)}~.
	\end{equation}
	with the production time and duration given by \cite{Sou:2021juh}
	\begin{equation}
		\qquad\qquad |k_I\tau_*|\approx0.66\mu-0.34\lambda\tilde\kappa~~,~~k_I\Delta \tau_*\approx1.4\sqrt{\mu}\left(1-\lambda\frac{0.32\tilde{\kappa}}{\mu}\right)~.
	\end{equation}
	We see that the particle production of transverse modes is suppressed (enhanced) for positive (negative) helicity, whereas the longitudinal mode is not affected by the chemical potential.	The vector boson propagators are
	\begin{eqnarray}
	&&D^{+-}_{i_1i_2}(\mathbf{k}_I,\tau,\tau')=\sum_{\lambda}\left[\epsilon^{\lambda}_{i_1}(-\hat{\mathbf{k}}_I)v^{\lambda}_{k_I}(\tau)\right]^*\epsilon^{\lambda}_{i_2}(-\hat{\mathbf{k}}_I)v^{\lambda}_{k_I}(\tau')~,\\
	&&D^{-+}_{i_1i_2}(\mathbf{k}_I,\tau,\tau')=\sum_{\lambda}\epsilon^{\lambda}_{i_1}(\hat{\mathbf{k}}_I)v^{\lambda}_{k_I}(\tau)\left[\epsilon^{\lambda}_{i_2}(\hat{\mathbf{k}}_I)v^{\lambda}_{k_I}(\tau')\right]^*~,\\
	&&D^{++}_{i_1i_2}(\mathbf{k}_I,\tau,\tau')={\theta}(\tau-\tau')D^{-+}_{i_1i_2}(\mathbf{k}_I,\tau,\tau')+\theta(\tau'-\tau)D^{+-}_{i_1i_2}(\mathbf{k}_I,\tau,\tau')~,\\
	&&D^{--}_{i_1i_2}(\mathbf{k}_I,\tau,\tau')={\theta}(\tau-\tau')D^{+-}_{i_1i_2}(\mathbf{k}_I,\tau,\tau')+\theta(\tau'-\tau)D^{-+}_{i_1i_2}(\mathbf{k}_I,\tau,\tau')~.
	\end{eqnarray}
	The polarization vector reads
	\begin{align}
	&\mathbf{\epsilon}^{\pm}(\hat{\mathbf{k}}_I)=\frac{1}{\sqrt{2(1-(\hat{\mathbf{n}}\cdot{\hat{\mathbf{k}}_I})^2)}}\left[(\hat{\mathbf{n}}-(\hat{\mathbf{n}}\cdot\hat{\mathbf{k}}_I)\hat{\mathbf{k}}_I)\pm i(\hat{\mathbf{k}}_I\times\hat{\mathbf{n}})\right]~,\nonumber\\
	&\mathbb{\epsilon}^{0}(\hat{\mathbf{k}}_I)=\hat{\mathbf{k}}_I~.
	\end{align}
	with $\hat{\mathbf{n}}$ being an arbitrary unit vector with $\hat{\mathbf{k}}_I\times{\hat{\mathbf{n}}}\neq 0$.
	
\begin{figure}[htp] 
		\centering
		\includegraphics[width=14.8cm]{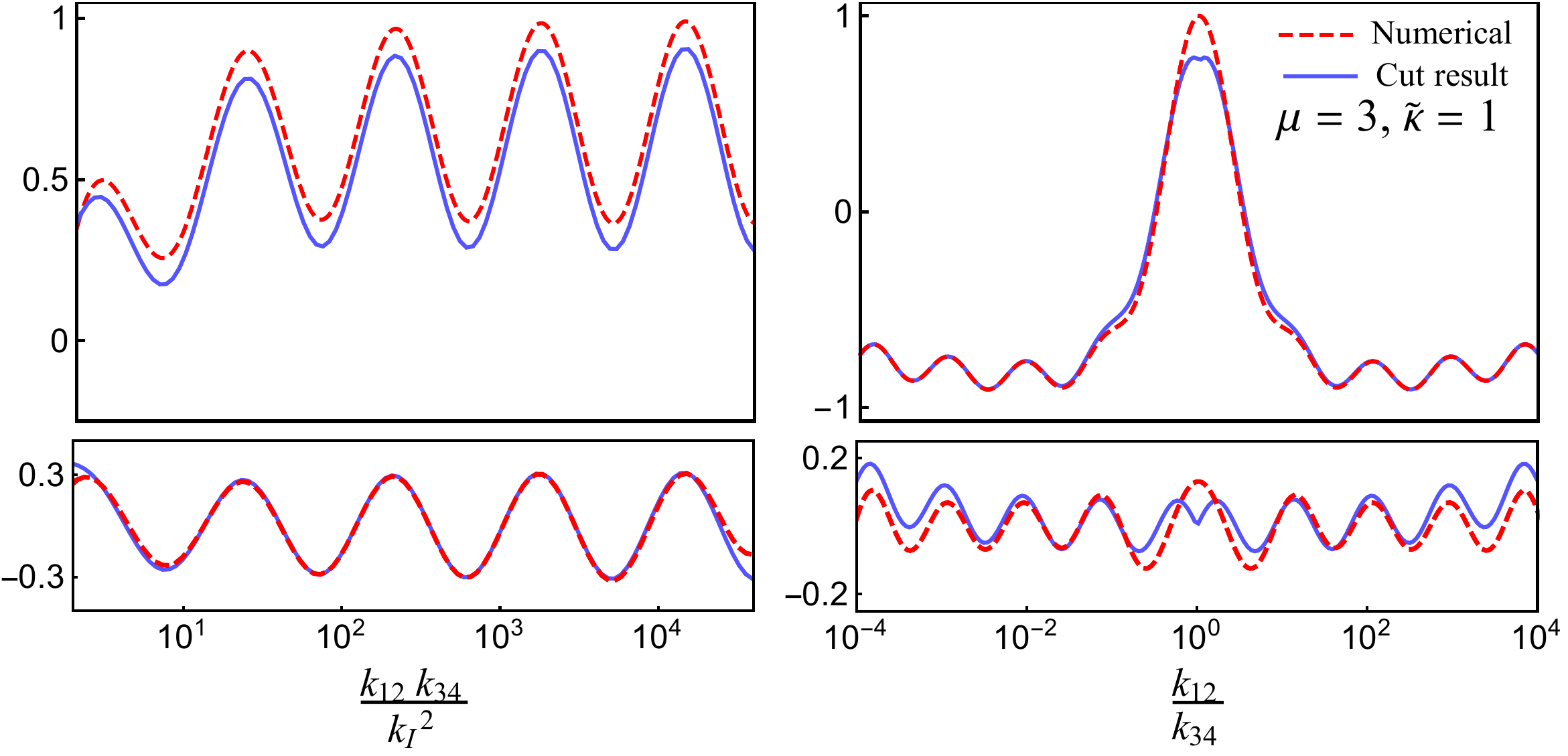}
		\caption{\label{vectormu3k1}The 4-point functions obtained from the cut rule (blue lines) and numerical integration (red dashed lines). \textit{Left panel}: Non-local type CC signals, where we have fixed $k_{12}/k_{34}=1$, and multiplied a factor $\left(k_{12}k_{34}/k_I^2\right)^{9/2}$ for better visualization. \textit{Right panel}: Local type CC signals, where we have fixed $k_{12}k_{34}/k_I^2=4\times10^4$. All results in the upper panels are normalized by their maxima. In the lower panels, we compare the CC signals filtered from the numerical result and that directly obtained via the cut.}
	\end{figure}

	After introducing the basic ingredients above, we now compute the 4-point function mediated by the vector boson. We assume the coupling
	\begin{align}
	\Delta\mathcal{L}_{int}=\lambda_3a \varphi'\partial_i\varphi A_i~.
	\end{align}
	For simplicity, we imagine that the external lines are distinguishable. One can always permute the external momentum after obtaining the result in which they are distinguishable. Following the cutting rule, as a first step, we focus on the TO part of the 4-point function,
	\begin{align}
	\nonumber{\langle\zeta_\mathbf{k_1}\zeta_\mathbf{k_2}\zeta_\mathbf{k_3}\zeta_\mathbf{k_4}\rangle}'_{s,\rm{TO}}=& \lambda_3^2\left(\frac{H}{\dot{\phi_0}}\right)^4\int_{-\infty}^{0}\int_{-\infty}^{0}d\tau d\tau'a(\tau)a(\tau')G_{++}(\mathbf{k_1},\tau,0)\partial_{\tau}G_{++}(\mathbf{k_2},\tau,0)\\ &\times k_1^{i_1}k_3^{i_2}D^{++}_{i_1 i_2}(\mathbf{k}_I,\tau,\tau')G_{++}(\mathbf{k_3},\tau',0)\partial_{\tau'}G_{++}(\mathbf{k_4},\tau',0)~.
	\end{align}
	The kinematic factor reads
	\begin{align}
	\nonumber\Pi^\lambda(\mathbf{k}_1,\mathbf{k}_3,\mathbf{k}_I)&\equiv\left(\mathbf{k}_1\cdot\mathbf{\epsilon}^{\lambda}(\hat{\mathbf{k}}_I)\right)\left(\mathbf{k}_3\cdot\mathbf{\epsilon}^{\lambda}(\hat{\mathbf{k}}_I)\right)^*\\
	&=
	\left\{
	\begin{aligned}
	&\frac{1}{2}\left[\mathbf{k}_1\cdot\mathbf{k}_3-(\mathbf{k}_1\cdot\hat{\mathbf{k}}_I)(\mathbf{k}_3\cdot\hat{\mathbf{k}}_I)-i\lambda \hat{\mathbf{k}}_I\cdot(\mathbf{k}_1\cross\mathbf{k}_3)\right]&\qquad\lambda=\pm~,\\
	&(\mathbf{k}_1\cdot\hat{\mathbf{k}}_I)(\mathbf{k}_3\cdot\hat{\mathbf{k}}_I)&\qquad\lambda=0~.
	\end{aligned}
	\right.
	\end{align}
	Step 2 being trivial, Step 3 gives the left/right blobs as
	\begin{eqnarray}
		\mathcal{F}_L(\tau)&=&\lambda_3 u_{k_1} u_{k_2}(0)a u^*_{k_1}u'^*_{k_2}(\tau)~,\\
		\mathcal{F}_R(\tau')&=&\lambda_3 u_{k_3} u_{k_4}(0)a u^*_{k_3}u'^*_{k_4}(\tau)~.
	\end{eqnarray}
	Step 4, cut the nested integral and discard the leftover commutator integral,
	\begin{eqnarray}
		{\langle\zeta^4\rangle}'_{s,TO}&=&S^>_{I}+\mathcal{I}_{I,com},
	\end{eqnarray}
	where
	\begin{align}
	S^>_{I}(k_{12},k_{34})=&\sum_{\lambda}\Pi^\lambda(\mathbf{k}_1,\mathbf{k}_3,\mathbf{k}_I)\left(\frac{H}{\dot{\phi_0}}\right)^4\int_{-\infty}^{0}d\tau\mathcal{F}_L v_{k_I}^\lambda(\tau)\int_{-\infty}^{0}d\tau'\mathcal{F}_R v^{\lambda*}_{k_I}(\tau')~,
	\end{align}
	and
	\begin{equation}
		\mathcal{I}_{I,com}=\sum_{\lambda}\Pi^\lambda(\mathbf{k}_1,\mathbf{k}_3,\mathbf{k}_I)\left(\frac{H}{\dot{\phi_0}}\right)^4\int_{-\infty}^{0}d\tau\int_{\tau}^{0}d\tau'\mathcal{F}_L(\tau) \mathcal{F}_R(\tau') \left(v^{\lambda*}_{k_I}(\tau)v^\lambda_{k_I}(\tau')-v^\lambda_{k_I}(\tau)v^{\lambda*}_{k_I}(\tau')\right)~.
	\end{equation}
	Again, $S^>_{I}$ can be analytically computed, we provide its expression in Appendix~\ref{AppExample3CCSignalExpression}~. Then in Step 5, we symmetrize the signal and obtain
	\begin{equation}
		S_I=\theta(k_{12}-k_{34})S_{I}^>(k_{12},k_{34})+\theta(k_{34}-k_{12})S_{I}^>(k_{34},k_{12})+(\text{c.c.}|_{\mathbf{k}_i\to-\mathbf{k}_i})~.\label{thirddiagsymtrS}
	\end{equation}
	Step 6 and 7, the EFT background is sourced at leading order by the local operator
	\begin{align}
		\Delta\mathcal{L}^{EFT}_I=\frac{\lambda_3^2}{\mu^2 H^2}\varphi'^2(\partial_i\varphi)^2~.
	\end{align}
	In the ``$s$''-channel, its contribution is
	\begin{equation}\label{thirddiagramEFT}
		B_I=\left(\frac{H}{\dot{\phi_0}}\right)^4\frac{\lambda_3^2H^6\mathbf{k_1}\cdot\mathbf{k_3}}{\mu^2} \left[\frac{4 k_1^2+(k_{1234}+4k_1) \left(k_{234}+3k_3\right)}{4k_1^3 k_2 k_3^3 k_4 k_{1234}^5}\right]~.
	\end{equation}
	Finally, we arrive at the result
	\begin{align}
		\langle\zeta_\mathbf{k_1}\zeta_\mathbf{k_2}\zeta_\mathbf{k_3}\zeta_\mathbf{k_4}\rangle_s'\approx S_{I}+B_I~,
	\end{align}
	with $S_I$ and $B_I$ given by (\ref{thirddiagsymtrS}) and (\ref{thirddiagramEFT}). We compare our cut result to numerics in Fig.~\ref{vectormu3k1}. Clearly, both the CC signals and the EFT background match the numerical result very well.
	
	
\begin{figure}[htp] 
		\centering
		\includegraphics[width=16cm]{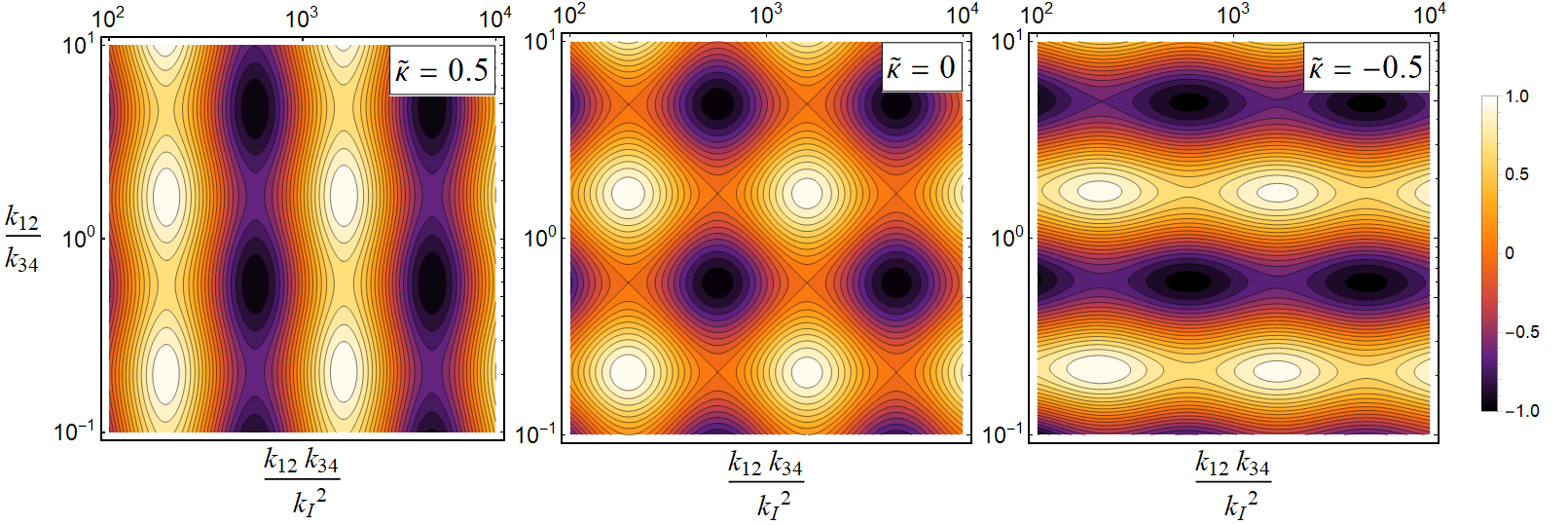}
		\caption{Signals of the negative helicity component, $(k_{12}k_{34}/k_I^2)^{9/2}S^>_{I}(k_{12},k_{34})|_{\lambda=-1}$, for different chemical potential choices ($\tilde\kappa=0.5, 0, -0.5$), with the mass fixed to be $\mu=3$. Each plot is normalized by their maximum. We see that oscillations in different directions dominates different parameter ranges. Without chemical potential, two CC signals share the same magnitude. As the chemical potential is turned on, the non-local type becomes dominant (subdominant) if the chemical potential is positive (negative).}\label{2d}
	\end{figure}

	To have a closer look at the signal strengths, we take the large-mass limit and the soft limit, then $S_I$ becomes
	\begin{align}
	\nonumber S_I\xrightarrow[k_I\to 0]{\mu\to\infty}~\propto\mu^3&\sum_{\lambda=-1}^1\Bigg\{e^{-\pi(\mu+\lambda\tilde\kappa)}\cos\left[\mu\ln\frac{4k_{12}k_{34}}{k_I^2}-\frac{\lambda^2\tilde\kappa^2}{\mu}\right]\times e^{i\lambda\phi_{(12,34)}}\\
	&~~~~~~~~~~~~-e^{-\pi\mu}\sin\left[\mu\ln\frac{k_{12}}{k_{34}}\right]\times\cos[\lambda\phi_{(12,34)}]\Bigg\}~.\label{thirddiagramlargemass}
	\end{align}
	We observe that both the amplitude and the phase of the non-local type CC signal is strongly dependent on the chemical potential $\tilde{\kappa}$. For instance, the non-local signal of the negative helicity mode $\lambda=-1$ is enhanced for $\tilde{\kappa}>0$, and suppressed for $\tilde{\kappa}<0$. In contrast, its local signal is chemical-potential independent, since it reflects the vertex production/decay process, whose amplitude only depends on the spacetime geometry. This is demonstrated in Fig.~\ref{2d} by plotting the negative-helicity signals on the projective plane spanned by $\frac{k_{12}k_{34}}{k_I^2}$ and $\frac{k_{12}}{k_{34}}$. To better visualize the oscillations, we have added a factor $(k_{12}k_{34}/k_I^2)^{9/2}$ and normalized each plots by their maximum. As we can see from the figure, there are two different oscillations along the horizontal and vertical directions, respectively. In the absence of chemical potential, two CC signals share the same magnitude. However, as soon as we turn on a non-zero chemical potential ($|\tilde\kappa|=0.5$), the non-local type becomes dominant (subdominant) if the chemical potential is positive (negative). The positive helicity mode behaves exactly in the opposite fashion, whereas the longitudinal mode is not affected at all. Observationally, the impact of chemical potential on CC signals of different helicities can be distinguished from their dihedral angle dependence. In fact, the case without chemical potential resembles linearly polarized light while that with chemical potential resembles elliptically polarized light. In this way, we can confirm that the two types of signals stem from drastically different phenomena. Notice also that, from the large-mass approximation (\ref{thirddiagramlargemass}), the parity-violating imaginary part of the 4-point function is only present in the non-local signal, not the local signal nor the EFT background. This agrees with the no-go theorem on parity-violating trispectra in single-field EFTs \cite{Liu:2019fag}.

   	\begin{figure}[htp] 
   	\centering
   	\includegraphics[width=10cm]{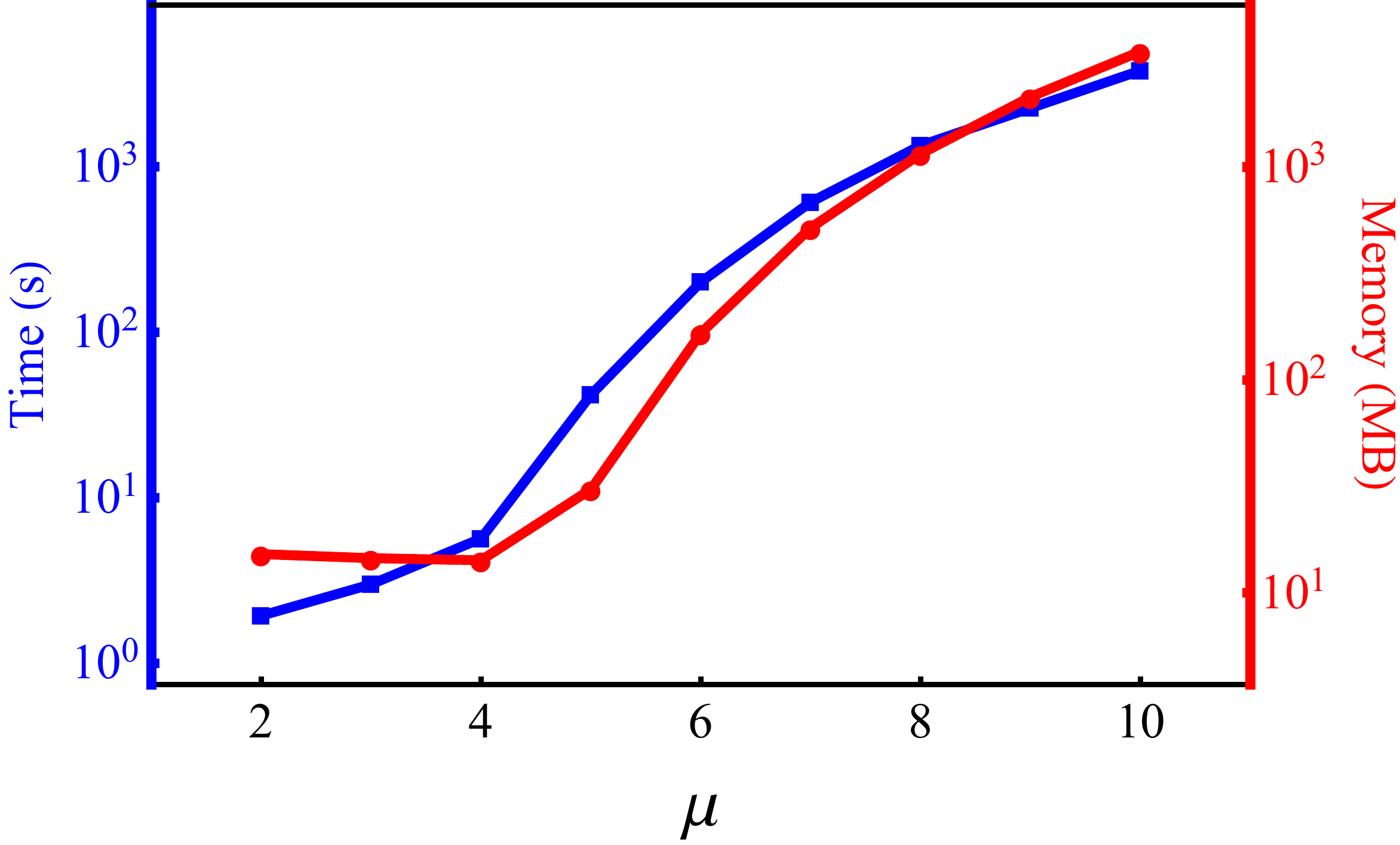}
   	\caption{The computational cost of a single data point in the configuration space using conventional numerics. The blue line and the red line show the time and memory consumption of computing Fig.~\ref{example3} with $k_{12}=k_{34}=20k_I$, where the relative numerical error is targeted at $\mathcal{O}(10^{-3})$. The chemical potential $\tilde{\kappa}$ is chosen as $\mu-2$~. As a comparison, the computational cost using the cut result is much lower, being around $\mathcal{O}(10^{-3})$s and $\mathcal{O}(10^{-2})$MB.}\label{timememory}
   \end{figure}
    Before concluding this section, we emphasize that with our cutting algorithm, the numerical efficiency is greatly enhanced. As mentioned in Sect.~\ref{Intro}, the conventional numerical integration approach to CC signals is extremely inefficient, and one has to balance between the integration precision and the computational cost. Take the diagram (Fig.~\ref{example3}) considered in this section as an example, we wish to monitor the computational cost of one data point in the configuration space, namely $(k_{12}=k_{34}=20k_I)$ for the non-local CC signal of the $\lambda=-1$ mode. In Fig.~\ref{timememory}, we plot the time (blue line) and memory (red line) consumption for different masses $\mu$, with the chemical potential chosen to be $\tilde{\kappa}=\mu-2$. Here we have targeted the relative numerical error at $\mathcal{O}(10^{-3})$. We see clearly that as mass grows larger, both the time and memory cost increase rapidly. For example, when $\mu=10,\tilde{\kappa}=8$, we need around one hour and approximately $3$GB of memory to compute even one data point. In contrast, with the help of the cutting rule, evaluating an analytical cut result only costs around $\mathcal{O}(1)$ms and $\mathcal{O}(10)$KB. This is an improvement of 5-6 orders of magnitude, which can be extremely helpful when sampling lots of data points in the trispectrum configuration space.

	\section{Summary and outlook}\label{SummarySect}
	Complicated by nested integrals of special functions, computing inflationary correlators and extracting oscillatory signals are difficult tasks in cosmological collider physics. In most cases, one can only rely on either numerics or work with approximate results of considerable systematic error. In this work, we leverage on the physical picture of bulk evolution, and to make a first step toward a systematic and efficient extraction method of CC signals. We started by a brief review of cosmological collider physics and the Schwinger-Keldysh formalism. After pointing out the technical difficulties for extracting CC signals, we moved on to a physical understanding of the bulk evolution of massive particles. We argued that at leading order, the bulk evolution history can be understood as a collection of resonance events. The oscillatory CC signals can be interpreted as the dynamical phases accumulated between these events, and are classified into two categories with distinct origins. Then, based on this resonance picture, we proposed a cutting rule for extracting analytical CC signals from a general tree diagram. The cutting algorithm is supplemented with three application examples, in which we find good agreement between the cut CC signals and those filtered from numerics. The computational efficiency is also found to be greatly improved compared to traditional numerical methods.
	
	As mentioned above, though this work attempts to understand the dynamics of massive fields during inflation and to develop efficient shortcuts to observables, there are certainly many improvements to make and future directions to explore. We conclude by listing some of them below.	
	\begin{enumerate}
		\item[$\bullet$] First, our cutting rule is subjected to two sources of errors. The EFT truncation error can sometimes be significant, especially when the particle mass is small or when the number of massive propagators is large. Therefore, we need to take into account the higher order terms in the large-mass EFT in a systematical way. Despite being tedious, this refinement is doable in principle. In contrast, the non-perturbative error at special momentum configurations may require new insights. We hope to reduce both errors in future works.
		\item[$\bullet$] Second, the specificity of the cutting algorithm makes it convenient to be implemented as a computer program. We will try to achieve this aim in the future. This may benefit model builders in cosmological collider who wish to skip the technicalities to direct signals. 
		\item[$\bullet$] Another limitation of our cutting rule is that, so far, it is only applicable to tree diagrams. At loop level, in addition to background pair production and vertex resonant production/decay of massive particles, more complex phenomena may happen, and the cutting rule may need further modifications.
		\item[$\bullet$] It is also interesting to study the relation between our phenomenological cutting rule to other more formal cutting rules. As mentioned in Sect.~\ref{CuttingRuleSummary}, they have crucial differences, but their shared features (most importantly, the factorization of non-local CC signals on the branch cut of the internal momentum) seem to hint a hidden connection between them. It is also helpful to understand our physical-picture-based cutting rule from a more fundamental perspective, which may aid to its generalization to other scenarios.
	\end{enumerate}
	
	\acknowledgments
	We would like to thank Zhong-Zhi Xianyu and Xingang Chen for helpful discussions. This work is supported in part by the National Key R\&D Program of China (2021YFC2203100), GRF Grants 16304418 and 16303819 from the Research Grants Council of Hong Kong, and the NSFC Excellent Young Scientist (EYS) Scheme (Hong Kong and Macau) Grant No. 12022516.

	\appendix
	\section{The commutator integral}\label{prcmint}
\begin{figure}[tp] 
		\centering
		\includegraphics[width=14cm]{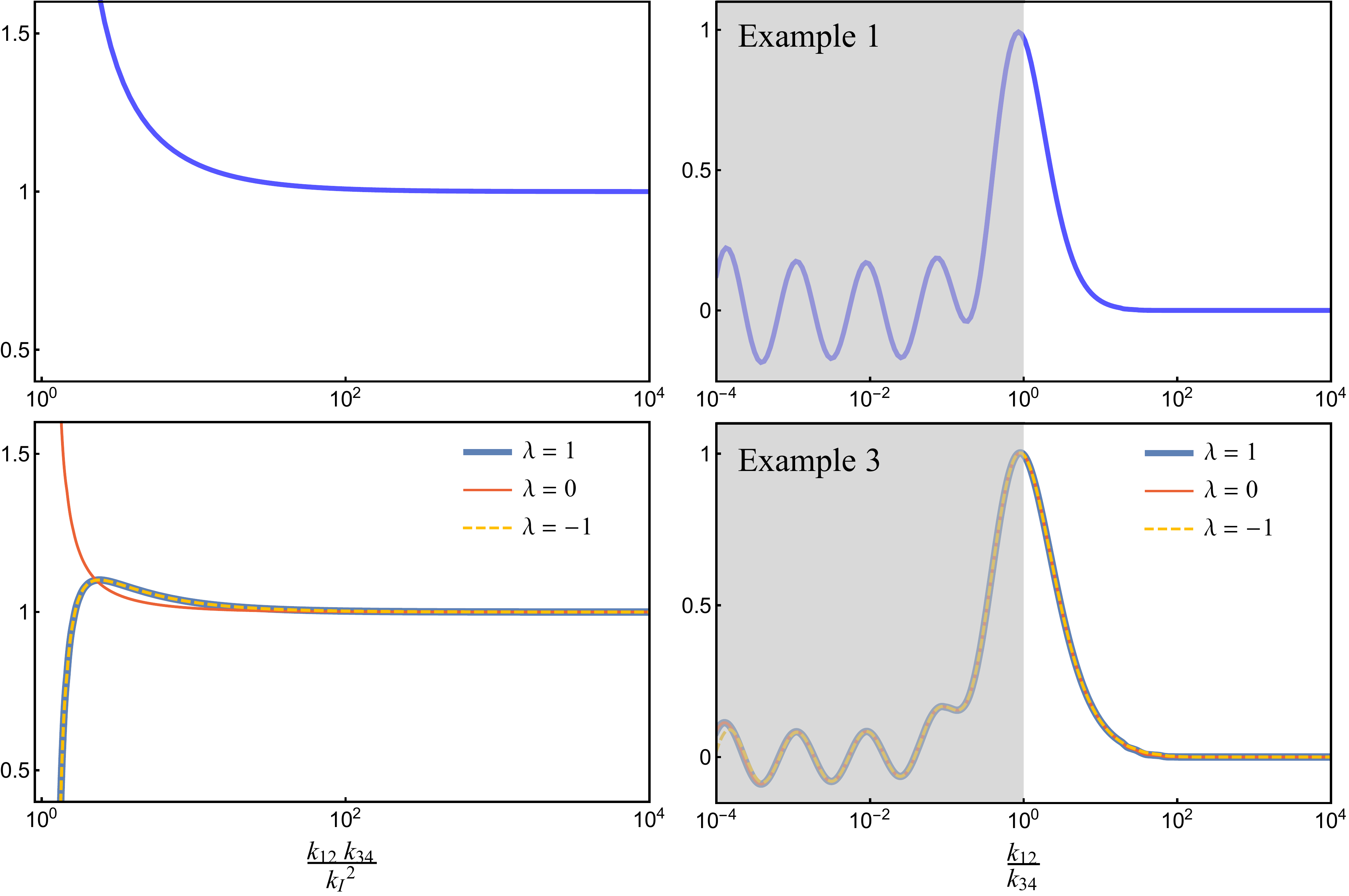}
		\caption{The numerical result of the commutator integral $\mathcal{I}_{I,com}$. \textit{First row}: The scalar exchange case in Example 1. \textit{Second row}: The vector exchange case in Example 3 with chemical potential. The left column shows the absence of non-local signals, where we have fixed $k_{12}/k_{34}=1$. $\mathcal{I}_{I,com}$ is multiplied by a factor $(k_{12}k_{34}/k_I^2)^{9/2}$ for better visualization, and is normalized by its value at $k_{12}k_{34}/k_I^2\gg 1$. The right column shows the absence of local signals in the region $k_{12}/k_{34}>1$. The region $k_{12}/k_{34}<1$ with oscillations is not our concern and is painted gray. Here we have fixed $k_{12}k_{34}/k_I^2=4\times10^4$ and normalized $\mathcal{I}_{I,com}$ by its maximum.}\label{timeorderno}
	\end{figure}

	In this appendix, we numerically verify that the commutator integrals in Example 1 and Example 3 contain neither non-local nor local signals. The result is shown in Fig.~\ref{timeorderno}. Clearly, the commutator integrals do not oscillate at all in the parameter region $k_{12}>k_{34}$. Thus it does not contain any CC signals, and can be safely deserted.

	\section{NTO part in Example 1}\label{intcc}
	In this appendix, we examine the NTO part of the diagram Fig.~\ref{example1}. With the coupling (\ref{inta}), the $s$-channel NTO part is
	\begin{align}
	\nonumber{\langle\zeta^4\rangle}_{s,\rm{NTO}}&=\tilde{c_3}^2u_{k_1} u_{k_2}u^*_{k_3}u^*_{k_4}(0)\int_{-\infty}^{0}d\tau a^2u'^{*}_{k_1}u'^{*}_{k_2}v^{*}_{k_I}\int_{-\infty}^{0}d\tau'a^2u'_{k_3}u'_{k_4}v_{k_I}+\text{c.c.}\\
	\nonumber&=\left(\frac{H}{\dot{\phi_0}}\right)^4\frac{c^2_3H^6e^{-\pi\mu}\pi }{16k_1k_2k_3k_4k_I^5}\\
	&\quad\times\left[\int_{0}^{\infty}dx_1{x_1}^{\frac{3}{2}}e^{-i\frac{k_{12}}{k_I}x_1}{H}^{(2)}_{-i\mu}(x_1)\int_{0}^{\infty}dx_2{x_2}^{\frac{3}{2}}e^{i\frac{k_{34}}{k_I}x_2}{H}^{(1)}_{i\mu}(x_2)+\text{c.c.}\right].
	\end{align}
	Using the Laplace transformation
	\begin{align}
	\nonumber\int_{0}^{\infty}dxx^nH^{(1)}_{i\mu}(x)e^{isx}&=\mathcal{L}\{x^n H^{(1)}_{i\mu}(x)\}(-is)\\
	&=\frac{i^ne^\frac{\pi\mu}{2}\big|\Gamma(1+n-i\mu)\big|^2}{\sqrt{\pi}2^n\Gamma\left(\frac{3}{2}+n\right)}{}_2F_1\Bigg[\begin{array}{c} 1+n-i\mu, 1+n+i\mu\\[2pt] \frac{3}{2}+n \end{array}\Bigg|\,\frac{1-s}{2}\Bigg]~,\label{HankelLaplace}
	\end{align}
	we can reduce the NTO CC signal to
	\begin{align}
	\nonumber{\langle\zeta^4\rangle}_{s,NTO}=&\left(\frac{H}{\dot{\phi_0}}\right)^4\frac{c^2_3H^6\pi^2(16\mu^4+40\mu^2+9)^2\sech^2(\pi\mu) }{2^{17}k_1k_2k_3k_4k_I^5}\\
	&\times{}_2F_1\Bigg[\begin{array}{c} \frac{5}{2}-i\mu, \frac{5}{2}+i\mu\\[2pt] 3 \end{array}\Bigg|\,\frac{k_I-k_{12}}{2k_I}\Bigg]\times{}_2F_1\Bigg[\begin{array}{c} \frac{5}{2}-i\mu, \frac{5}{2}+i\mu\\[2pt] 3 \end{array}\Bigg|\,\frac{k_I-k_{34}}{2k_I}\Bigg]+\text{c.c.}~.
	\end{align}
	To see the Boltzmann suppression factor more explicitly, we take the soft limit of internal momentum as well as the large-mass expansion,
	\begin{eqnarray}
	{\langle\zeta^4\rangle}_{s,NTO}\propto e^{-2 \pi  \mu } \mu ^3 \left\{\sin \left[\mu  \log \left(\frac{4k_{12}k_{34}}{k_I^2}\right)\right]+\cos \left[\mu  \log
	\left(\frac{k_{34}}{k_{12}}\right)\right]\right\}~.
	\end{eqnarray}
	As a result of the smallness of the Boltzmann factor $e^{-2 \pi  \mu }$, the NTO part is always subdominant compared to the TO part, thus we can safely neglect it at leading order.

	\section{Expressions of CC signals in Example 2}\label{AppExample2CCSignalExpression}
	Using the Laplace transformation for Hankel functions (\ref{HankelLaplace}), we can obtain the analytical form of CC signals as
	\begin{align}
		\nonumber S^>_{(12,34)}(k_{12},k_{34})=&-\left(\frac{H}{\dot{\phi_0}}\right)^4\frac{\lambda^2\rho^4\pi^2\left(16\mu^4+40\mu^2+9\right)^2\sech(\pi\mu)^2}{2^{19}k_1k_2k_3k_4k_I^5\nu^8H^2}\\
		&\times{}_2F_1\Bigg[\begin{array}{c} \frac{5}{2}-i\mu, \frac{5}{2}+i\mu\\[2pt] 3 \end{array}\Bigg|\,\frac{k_I+k_{12}}{2k_I}\Bigg]\times{}_2F_1\Bigg[\begin{array}{c} \frac{5}{2}-i\mu, \frac{5}{2}+i\mu\\[2pt] 3 \end{array}\Bigg|\,\frac{k_I-k_{34}}{2k_I}\Bigg]
	\end{align}
	in the $(12,34)$ channel and
	\begin{align}
	    S^>_{(1,234)}(k_{1},k_{234})&=-\left(\frac{H}{\dot{\phi_0}}\right)^4\frac{\lambda^2\rho^4\pi\sech(\pi\nu)\bigg|\Gamma\left(\frac{9}{2}-i\nu\right)\bigg|^2}{12288\mu^2\nu^6H^2k_1^6k_2k_3k_4}{}_2F_1\Bigg[\begin{array}{c} \frac{9}{2}-i\nu, \frac{9}{2}+i\nu\\[2pt] 5 \end{array}\Bigg|\,\frac{k_{1234}}{2k_1}\Bigg]
	\end{align}
	in the $(1,234)$ channel. Signals in the other three channels ($(2,134)$, etc.) are obtained via simple replacements.
	
	\section{Expressions of CC signals in Example 3}\label{AppExample3CCSignalExpression}
	The analytical expression of CC signals in Example 3 can be computed by noticing the Laplace transformation of the Whittaker function,
	\begin{align}
		\int_{0}^{\infty}x^n W_{i\lambda \tilde\kappa,i\mu}(c x)e^{-i s x}&=\mathcal{L}\{x^n W_{i\lambda \tilde\kappa,i\mu}(c x)\}(is) \nonumber\\
		&=c^{-1-n}\bigg|\Gamma\left(\frac{3}{2}+n-i\mu\right)\bigg|^2{}_2\mathbf {F}_1\Bigg[\begin{array}{c} \frac{3}{2}+n-i\mu, \frac{3}{2}+n+i\mu\\[2pt] 2+n-i\kappa\lambda \end{array}\Bigg|\,\frac{1}{2}-\frac{is}{c}\Bigg]~.
	\end{align}
	Here ${}_2\mathbf {F}_1$ is the regularized hypergeometric function
	\begin{align}
		{}_2\mathbf {F}_1\Bigg[\begin{array}{c} a,b\\[2pt] c \end{array}\Bigg|\,z\Bigg]\equiv
		{}_2{F}_1\Bigg[\begin{array}{c} a,b\\[2pt] c \end{array}\Bigg|\,z\Bigg]/\Gamma(c)~.
	\end{align} 
	Denoting the factorized time integrals for transverse modes as
	\begin{align}
		\nonumber\mathcal{I}^\lambda_L&\equiv\int_{0}^{\infty}\left(1+i\frac{k_1}{k_I}x\right)W_{i\lambda \tilde\kappa,i\mu}e^{-i\frac{k_{12}}{k_I}x}(-2ix)dx\\
		&={i\pi(4\mu^2+1)\sech(\pi\mu)}\nonumber\\
		&\times\frac{1}{8}\left\{{}_2\mathbf {F}_1\Bigg[\begin{array}{c} \frac{3}{2}-i\mu,\frac{3}{2}+i\mu\\[2pt] 2-i\lambda\tilde{\kappa} \end{array}\Bigg|\,\frac{k_I+k_{12}}{2k_I}\Bigg]-\frac{\left(\mu^2+\frac{9}{4}\right)k_1}{2k_I}{}_2\mathbf {F}_1\Bigg[\begin{array}{c} \frac{5}{2}-i\mu,\frac{5}{2}+i\mu\\[2pt] 3-i\lambda\tilde{\kappa} \end{array}\Bigg|\,\frac{k_I+k_{12}}{2k_I}\Bigg]\right\}~,\\
		\nonumber\mathcal{I}^\lambda_R&\equiv\int_{0}^{\infty}\left(1+i\frac{k_3}{k_I}x\right)W_{-i\lambda \tilde\kappa,-i\mu}e^{-i\frac{k_{34}}{k_I}x}(2ix)dx\\
		&=\mathcal{I}^\lambda_L\bigg|_{k_{1,2}\rightarrow-k_{3,4}}^*~,
	\end{align}
	while for the longitudinal mode
	\begin{align}
		\nonumber\mathcal{I}^0_L&\equiv\int_{0}^{\infty}\left(1+i\frac{k_1}{k_I}x\right)\left(\frac{-2xW_{0 ,1+i\mu}(-2ix)-(2\mu+i)W_{0,i\mu}(-2ix)}{\sqrt{1+4\mu^2}}\right)e^{-i\frac{k_{12}}{k_I}x}dx\\
		&=\frac{\pi\sech(\pi\mu)}{16(k_{12}-k_I)^2(k_{12}+k_I)(1+4\mu^2)^\frac{1}{2}}\nonumber\\
		&\times\left\{c^{(1)}{}_2F_1\Bigg[\begin{array}{c} \frac{3}{2}-i\mu, \frac{3}{2}+i\mu\\[2pt] 2 \end{array}\Bigg|\,\frac{k_I+k_{12}}{2k_I}\Bigg]+c^{(2)}{}_2F_1\Bigg[\begin{array}{c} \frac{5}{2}-i\mu, \frac{3}{2}+i\mu\\[2pt] 2 \end{array}\Bigg|\,\frac{k_I+k_{12}}{2k_I}\Bigg]\right\}~,\\
		\mathcal{I}^0_R&=\mathcal{I}^0_L\bigg|_{k_{1,2}\rightarrow-k_{3,4}}^*~,
	\end{align}
	with
	\begin{align}
		c^{(1)}&=(1+\mu^2)\left\{2k_I^2(-4k_I+k_2(9-6i\mu))+k_{12}k_I\left(4k_I(2i\mu-5)-k_2(4\mu^2+9)\right)\nonumber\right.\\
		&\left.+k_{12}^2\left(-3k_2+17k_I-4ik_2\mu+4(k_I-k_2)\mu^2\right)+k_{12}^3(4\mu^2+8i\mu+5)\right\}~,\\
		c^{(2)}&=2\left(3-2i\mu(1+6i\mu+4\mu^2)\right)\left(k_{12}^3+k_{12}^2k_1+(2k_1-k_2)k_I^2\right)~,
	\end{align}
	we obtain the analytical result of CC signals in the diagram Fig.~\ref{example3},
	\begin{align}
		S^>_{I}(k_{12},k_{34})=&\sum_{\lambda}\Pi^\lambda(\mathbf{k}_1,\mathbf{k}_3,\mathbf{k}_I)\left(\frac{H}{\dot{\phi_0}}\right)^4\frac{\lambda^2_3H^6e^{-\pi\lambda\tilde{\kappa}}}{2^5k_1^3k_2k_3^3k_4k_I^3}\mathcal{I}^{\lambda}_L\mathcal{I}^{\lambda}_R~.
	\end{align}

\bibliographystyle{utphys}
\bibliography{ref}


%
%
%
%
%
%
%
%
\end{document}